\documentclass[12pt, reqno, a4paper]{amsart}
\usepackage{graphicx}
\usepackage{fancyhdr}
\usepackage{enumerate}
\usepackage{amsmath}

\usepackage{mwe}
\usepackage{subcaption}

\usepackage{amsthm}
\usepackage{mathabx}
\usepackage{amssymb}
\usepackage{relsize}
\usepackage{pdfsync}
\usepackage[colorlinks=true,linkcolor=blue,citecolor=magenta]{hyperref}

\usepackage[normalem]{ulem}

\usepackage{xcolor}

\usepackage{cancel}

\usepackage{calligra}

\usepackage{tikz}

\usepackage{changebar}
\usepackage{pdfsync}
\usetikzlibrary{decorations.pathmorphing,patterns}
\usetikzlibrary{calc,arrows}

\usepackage{hyperref}

\usepackage{cleveref}

\usepackage{pgf}

\usetikzlibrary{automata}
\usetikzlibrary{positioning}

\tikzset{
    state/.style={
           rectangle,
           rounded corners,
           draw=black, very thick,
           minimum height=2em,
           inner sep=2pt,
           text centered,
           },
}

\usepackage{MnSymbol}

\usepackage{xcolor}

\usepackage{changebar}

\usepackage[normalem]{ulem}

\usepackage{mathtools}

\makeindex
\newcommand{\cal}{\mathcal}

\newcommand\bbR{{\mathbb R}}
\newcommand\bbZ{{\mathbb Z}}

\newcommand\lang{\langle \langle}
\newcommand\rang{\rangle \rangle}

\newcommand\bbE{{\mathbb E}}

\newcommand\R{{\mathbb R}}

\newcommand\bbI{{\mathbb I}}

\newcommand\eps{\epsilon}

\newcommand \ga{\gamma}
\newcommand \om{\omega}

\newcommand\Om{\Omega}

\renewcommand{\le}{\leqslant}
\newcommand{\dd  }{\mathrm{d}}

\renewcommand{\tilde}{\widetilde}
\renewcommand{\bar}{\overline}
\numberwithin{equation}{section}

\newcommand{\cF}{{\mathcal F}} 

\begin{document}

\title[Periodically forced harmonic chain]{On the behaviour
  of a periodically forced and thermostatted harmonic chain}

\author{Pedro L. Garrido}
 \address{Pedro L. Garrido\\Universidad de Granada\\Granada Spain} 
\email{{\tt garrido@onsager.ugr.es}}

 \author{Tomasz Komorowski}
 \address{Tomasz Komorowski, Institute of Mathematics,
   Polish Academy Of Sciences, Warsaw, Poland\\
   \and Institute of Mathematics, Maria Curie-Sk\l odowska University,
   Lublin, Poland.} 
\email{{\tt tkomorowski@impan.pl}}

\author{Joel L. Lebowitz}
\address{Joel L. Lebowitz, Departments of Mathematics and Physics,  Rutgers University}
\email{\tt lebowitz@math.rutgers.edu}

 \author{Stefano Olla}
 \address{Stefano Olla, CEREMADE,
   Universit\'e Paris-Dauphine, PSL Research University \\
 \and Institut Universitaire de France\\ \and
GSSI, L'Aquila}
  \email{\tt olla@ceremade.dauphine.fr}

\date{\today {\bf File: {\jobname}.tex.}}

\begin{abstract}
  We consider a   chain consisting of $n+1$  pinned harmonic oscillators
subjected on
the right to a time dependent periodic force $\cF(t)$ while 
  Langevin thermostats are attached at both endpoints of the chain.
We show that for long times the system is described by a Gaussian
measure whose covariance function is independent of the force, while
the means are periodic. We compute explicitly the work and energy due
to the periodic force for all $n$ including $n\to\infty$. 
\end{abstract}

\thanks{We thank David Huse for helpful discussions, and the IAS for the
  hospitality during part of this work. P.G.  acknowledges the support of 
  the Project I+D+i Ref.No.PID2020-113681GB-I00,
  financed by MICIN/AEI/10.13039/501100011033 and FEDER “A waytomakeEurope.”
T.K. acknowledges the support of the NCN grant 2020/37/B/ST1/00426.} 

\maketitle


\section{Introduction}
\label{intro}

In this work we consider the conversion of work into heat in a simple
model system: a pinned harmonic chain of $n+1$ particles on which 
work is  {performed}  by an external periodic force  {acting at one of
the endpoints.}
The system is also in contact with thermal reservoirs,  {placed at both
of its endpoints}, which absorb the energy generated by the work. In
the absence of the reservoirs the response of the system to the external forcing depends entirely on whether the frequency $\om$ of the external force coincides with the normal frequencies of the
chain {$\{ \om_j, j=0,\ldots, n\}$}. When $\om\neq \om_j$
the system adjusts itself to be out of phase with the force
so that there is no work done on the average.
If on the other hand the system is in resonance
with the force, i.e. $\om=\om_j$ for some $j$, then the amplitude of
the oscillation tends to infinity 
as time $t\to \infty$. 

The situation is different in the presence of the thermostats.
They  cause the oscillations at resonance to be damped and  {as a
  result} the work done by the force is strictly finite for all values of $\om$.

There is still a strong dependence on $\om$,  {as far as  the magnitude
of the work is concerned, when $n$ gets large.}
This difference becomes qualitative when $n\to\infty$ and the spectrum of the harmonic chain
becomes dense in an interval $\cal I$. The work done and the internal energy of the chain
depend strongly on whether $\om$ lies in the interior of $\cal I$, or not.

Due to the linearity of the system there is a clear division, in the
long time properties of the system, between those due to temperatures
of the thermal reservoirs  and those due to the external force.
 {The energy flowing through the system {as a result of} the presence of the thermal reservoirs
  we call \emph{thermal energy}. It is not influenced by the external force and its behavior
   is the same as in \cite{RLL67} and \cite{Nak70}.
  The energy flow due to the work of the external force we call \emph{mechanical energy}.
  It is independent of the temperatures of the reservoirs,
  and it is influenced only by the corresponding damping.
  For finite $n$ and pure damping equal on both sites this
  was computed in section 4 of \cite{prem}
  in terms of the Green function of the corresponding damped harmonic chain.
  The main objective of the present note is the exact calculation of the asymptotic behavior
as $n\to\infty$ of the work and the mechanical energy.}
Calculation of these quantities, turns out to be  quite complicated,
but  leads to explicit expressions for  their
asymptotics.
{In particular we show that, for forcing frequency outside $\cal I$, the work,
  the mechanical energy and its flow become negligible as $n\to \infty$. Inside $\cal I$
  these quantities oscillate fast and their asymptotic behavior can be described
  in terms of Young measures.}

The results of the present work remain also valid in the case of
unpinned harmonic chain. It suffices  to  set the pinning constant
$\om_0=0$ in our formulas describing the work and energy
functionals. {Obviously we now always have \linebreak$\om>\om_0=0$
  and consider the motion relative to the center of mass positioned at zero.}

For anharmonic interaction the situation is qualitatively very
different. The non-linearity produces many new effects described in \cite{prem}.
We have also studied   the case of a harmonic chain with a random velocity flip \cite{klo1}.

\section{Description of the System}
\label{sec:description-system}

The configurations of our system, consisting of $n+1$  pinned harmonic oscillators, are described by
\begin{equation}
  \label{eq:1}
  (\mathbf q, \mathbf p) =
  (q_0, \dots, q_n, p_0, \dots, p_n) \in \Omega_n:=\R^{n+1}\times\R^{n+1}. 
\end{equation}
We should think {of} the positions $q_x$ as the relative displacement of an
atom from a point
 $x$ belonging to the integer lattice interval $ {\bbI_n=
   \{0,\ldots,n\}}$ and $p_x$ as  its respective momentum.

The total energy of the chain is given by the Hamiltonian:
\begin{equation}
\label{Hn}
\mathcal{H}_n (\mathbf q, \mathbf p):=
\sum_{x=0}^n e_x (\mathbf q, \mathbf p),
\end{equation}
where the microscopic energy density at $x$ is given by
\begin{equation}
\label{Ex}
e_x (\mathbf q, \mathbf p):=  \frac{p_x^2}2 +
\frac12 (q_{x}-q_{x-1})^2 +\frac{\om_0^2 q_x^2}{2},\quad x \in\bbI_n .
\end{equation}
Here we let $q_{-1}:=q_0$.

The  microscopic dynamics of
the process  
describing the total chain is  given   by
\begin{equation} 
\label{eq:flip}
\begin{aligned}
  \dot   q_x(t) &= p_x(t) ,
 \qquad \qquad x\in \bbI_n,\\
   \dot    p_x(t) &=  \Delta_x q_x-\om_0^2 q_x  ,\quad
  x\in \bbI_n^o= \{1, \dots, n-1\}
  \end{aligned} \end{equation}
and at the boundaries by
\begin{align}
     \dd   p_0(t) &=   \; \Big(q_1(t)-q_0(t) - \om_0^2 q_0 \Big) \dd   t
                    -
                    2  \gamma_- p_0(t) \dd t
                    +\sqrt{4  \gamma_- T_-} \dd \tilde w_-(t),
                    \vphantom{\Big(} \label{eq:rbdf}\\
  \dd   p_n(t) &=  \; \Big(q_{n-1}(t) -q_n(t) -\om_0^2 q_n(t) \Big)  \dd   t  +
                 \; \cF(t/\theta)  \dd t  -
                    2   \gamma_+ p_n(t) \dd t
                    +\sqrt{4  \gamma_+ T_+} \dd \tilde w_+(t).
                     \vphantom{\Big(} \notag
\end{align}
Here   $\Delta q_x=q_{x+1}+q_{x-1}-2q_x$, $x\in\bbZ$ is the  laplacian
on the integer lattice $\bbZ$, $\om_0>0$ is a pinning constant,
$\tilde w_-(t)$ and $\tilde w_+(t)$ are two independent standard one dimensional Wiener
processes 
and $\gamma_\pm$ are non-negative constants that 
describe   the respective strengths of the Langevin thermostats.

We assume the force $ \cF(t)$ to be a smooth periodic function
of period $1$ and parameter $\theta$ rescales the period.  We will
suppose, without losing generality, that
\begin{equation}
  \label{eq:2}
  \int_0^1  \cF(t) \dd t = 0, \qquad  \int_0^1  \cF(t)^2 \dd t > 0.
\end{equation}

The generator of the dynamics is given by
\begin{equation}
  \label{eq:7}
  \mathcal G_t =  \mathcal A_t   
  + 2\sum_{\iota\in\{-,+\}}\gamma_{\iota} S_\iota,
\end{equation}
where
\begin{equation}
  \label{eq:8}
  \mathcal A_t = \sum_{x=0}^n p_x \partial_{q_x}
  + \sum_{x=0}^n  (\Delta q_{x}-\om^2_0q_x) \partial_{p_x}
  + \cF(t/\theta)  \partial_{p_n}.
\end{equation}
By convention we let $q_{n+1}:=q_n$ and {$q_{-1}= q_0$}.
Furthermore
 \begin{equation}
   \label{eq:10}
   S_-= T_- \partial_{p_0}^2 - p_0 \partial_{p_0},\qquad  S_+= T_+ \partial_{p_n}^2 - p_n \partial_{p_n}
 \end{equation}

The energy currents are  
\begin{equation}
\label{eq:current}
\begin{aligned}
  & \mathcal G_t e_x  = j_{x-1,x} - j_{x,x+1} ,\\
 & j_{x,x+1}:=- p_x (q_{x+1}- q_x) , \qquad \text{if } x \in \{0,...,n-1\}, 
\end{aligned}
\end{equation} 
and at the boundaries 
  \begin{equation} \label{eq:current-bound}
     j_{-1,0} := 2 { \gamma}_- \left(T_- - p_0^2 \right)
      \qquad
  j_{n,n+1} :=   -2 { \gamma}_+ \left(T_+ - p_n^2 \right) - \cF(t/\theta)  p_n.
\end{equation}

We are interested in the long time behavior of the system.
In the absence of the external forcing, $\cF(t)\equiv0$, this is just
the model considered in \cite{RLL67}, with $\om_0=0$, and in
\cite{Nak70} for $\om_0>0$.
In the case when $\cF(t)\equiv0$, starting
with any initial configuration  $(\mathbf q(0), \mathbf p(0))$ (or any
initial probability distribution $\mu_0(\dd\mathbf q, \dd\mathbf p)$)
the system approaches a stationary Gaussian distribution
$\mu_{\rm stat}(\dd\mathbf q, \dd\mathbf p)$, in which the expectation values
of  $q_x$ and $p_x$ vanish, i.e. $\bar q_x(t)=0$ and $\bar p_x(t)=0$, while the covariances between components
of $(\mathbf q, \mathbf p)$ are given explicitly.

In particular the
expectation of the energy current $\bar j_{x,x+1}$ between sites $x$
and $x+1$, that is independent of $x$ and $t$, is given by
\begin{equation}
  \label{012905-23}
  \bar j_{x,x+1}=(c+o(1))(T_--T_+),\quad \mbox{as }n\gg1,
\end{equation}
with
$$
c=\frac{\ga}{1+4\ga^2+2\ga\om_0(\ga\om_0+\sqrt{1+4\ga^2+ (\ga\om_0)^2})},
$$
when $\ga_-=\ga_+=\ga$, see \cite[formula (37), p. 240]{Nak70}. In the
case $\om_0=0$ the term $o(1)$ in the  formula \eqref{012905-23} can be
omitted (no dependence on $n$) and we have $c=\frac{\ga}{1+4\ga^2}$, see \cite[formula (40),
p. 241]{Nak70}.

Eq.  \eqref{012905-23} implies that the thermal conductivity is proportional to $n$
- the size of the system - and becomes
infinite in the limit  $n\to+\infty$, see also \cite{RLL67}. In fact
the ''temperature''  {$T_x$,  defined as the variance of $p_x^2$,}   is independent of $x$, except near the boundary points
$x=0,n$. Adding now the periodic force  of period $\theta$   leads, as
$t\to+\infty$, to a Gaussian, periodic  stationary state $\{\mu_t^P, t\in[0,+\infty)\}$,
whose covariances are 
 {the same as in the case when no force is applied.
For any functions $F=F(\bf q,\bf p)$ and $G=G(t)$ define
  \begin{equation}
    \label{eq:9}
    \bar F(t) = \int_{\Om_n} F \dd \mu_t^P\quad \mbox{and}\quad
    \lang G\rang= \frac{1}{\theta}\int_0^\theta G(t) \dd t.
  \end{equation}
  The periodic  stationary state has the property that $\lang\bar{\mathcal G  F}\rang = 0$
  for any $F$ in the domain of $\mathcal G_t$. 
}

The
expectation values of the position and momentum $\bar q_x(t)$ and
$\bar  p_x(t) $ are now
$\theta$-periodic and independent of the temperature of the
reservoirs. They are given by
$$
\left(
  \begin{array}{c}\bar{\bf q}(t)\\
    \bar{\bf p}(t)
    \end{array}\right)=\int_{-\infty}^t
e^{-A(t-s)}\;\cF(s/\theta){\rm e}_{p,n+1}\dd s.
$$
Here $A$ is a $2\times 2$ block matrix made of $(n+1)\times (n+1)$
matrices of the form
$$
A=
\left(
  \begin{array}{cc}
    0&-{\rm Id}_{n+1}\\
    -\Delta_{\rm N}+\om_0^2& \Gamma
  \end{array}
\right),
$$
where ${\rm Id}_{n+1}$ is  the $(n+1)\times (n+1)$ identity matrix,
$\Delta_{\rm N}$ is the Neumann laplacian
  on $\bbI_n$:
\begin{equation}
  \label{Delta-N}
  \Delta_{\rm N} f_x:= \Delta f_x,\quad x\in\bbI_n^o\quad\mbox{and }\quad
  \Delta_{\rm N} f_0=f_1-f_0,\quad  \Delta_{\rm N} f_n=f_{n-1}-f_n.
\end{equation}
Furthermore $\Gamma$ is the diagonal matrix
$$
\Gamma=2\left(
  \begin{array}{ccccc}
    \gamma_-&0&\ldots&0&0\\
    0& 0&\ldots&0&0\\
    \vdots& \vdots&\vdots&\vdots&\vdots\\
    0& 0&\ldots&0&0\\
     0& 0&\ldots&0&\ga_+
  \end{array}
\right) .
$$
The   column vector ${\rm e}_{p,n+1}$
is given by
$
{\rm e}_{p,n+1}^T=[\underbrace{0,\ldots,0}_{2n+1 - \rm{times}},1].
$
 {
  Notice that the first of the conditions \eqref{eq:flip} implies that
  $\lang \bar p_x\rang = 0$, while the second gives $\lang \bar q_x\rang= 0$.}

The expected value of  energy, averaged over a period,
 breaks up into the mechanical part, 
coming from the averaged position $\bar{\bf q}(t)$  and momentum
$\bar{\bf p}(t)$, which is independent of the temperature of the
reservoirs, and the thermal part, which is independent of the external force. More
precisely
\begin{equation}
  \label{e-thm}
\lang e_x\rang=\lang e_x^{\rm mech}\rang+\lang e_x^{\rm th}\rang,
\end{equation}
where the mechanical component of the energy is given by
\begin{equation}
  \label{e-kin}
e_x^{\rm mech}(t):=\frac12\Big[\bar p_x^2(t)+\om_0^2
\bar q_x^2(t)+\big(\bar q_x(t)-\bar q_{x-1}(t)\big)^2\Big],\quad x\in\bbI_n.
\end{equation}
and the thermal part is
\begin{equation}
  \label{e-thm1}
e_x^{\rm th}(t):=\frac12\bbE\Big[ p_x'(t)^2+\om_0^2
q_x'(t)^2+\big(q_x'(t)- q_{x-1}'(t)\big)^2\Big],\quad x\in\bbI_n.
\end{equation}
where $q_x'(t)=q_x(t)-\bar q_x(t)$ and $p_x'(t) =p_x(t)-\bar p_x(t)$
and $\bbE$ denotes the average with respect to the initial data and
the realizations of the Wiener processes in \eqref{eq:rbdf}. {As
  before, we adopt the convention $\bar q_{-1}(t):=\bar q_{0}(t)$ and
  likewise $ q_{-1}'(t):= q_{0}'(t)$.}

As already mentioned in the Introduction one of the  goals of
the present paper is  to  describe  the work done by the force on the
system. It is given by
\begin{equation}
  \label{013105-23}
W(n)=\frac{1}{\theta}\int_0^\theta \cF(t/\theta)
\bar p_n(t) \dd t.
\end{equation}
$W(n)$ is always positive, generates energy fluxes into
the two heat reservoirs. Furthermore, we   describe the
time average of the 
mechanical energy functional given by eq. \eqref{e-kin}. Its
thermal counterpart does not depend on  time and has been described  in
 \cite{Nak70,RLL67}.  {We mention here also that the case $n=0$, i.e. a
 single  oscillator in contact with a heat bath and driven by an
 external unbiased time-periodic force, has been fully characterized in \cite{Yag2017}. }

\section{Results}

 {In what follows we will  use the dispersion relation of the
  infinite chain given by
  \begin{equation}
       \label{032806-23}
       \om(r)=\sqrt{\om_0^2+4\sin^2\left(\frac{\pi r}{2}\right)}, \qquad r\in[0,1]
     \end{equation}
     and its inverse defined for $\omega\in \cal I := [\omega_0,
     \sqrt{\omega_0^2 +4}]$ by the formula
     \begin{equation}
       \label{eq:11}
       r(\omega) =\frac 2\pi \arcsin \left(\frac 12 \sqrt{\omega^2 - \omega_0^2}\right), 
     \end{equation}
}

\subsection{Work done by the force on the system}

\label{work}


The work $W(n)$ performed by the force on the system, see
\eqref{013105-23},   depends on the period $\theta$.
Considering for simplicity the simple mode case when
\begin{equation}
  \label{022905-23}
\cF(t/\theta )=F\cos(\om t),\quad \om:=\frac{2\pi }{\theta}
\end{equation}
the work  done  is given by (see 
Appendix):
\begin{equation}
  \label{033105-23}
W(\om,n)=\big(\om F\big)^2 \frac{N(\om,n)}{D(\om,n)}.
\end{equation}
Here
\begin{align}
  \label{020206-23}
  N(\om,n)=\ &\gamma_-G^{1}(\om,n)^2+\gamma_+G^{0}(\om,n)^2
            +4{\gamma_-^2\gamma_+}\om^2\left(G^{0}(\om,n)^2-G^{1}(\om,n)^2\right)^2\nonumber\\
  D(\om,n)=\ &1+8\gamma_-\gamma_+ {\om^2}G^1(\om,n)^2
            +4\om^2G^0(\om,n)^2(\gamma_-^2+\gamma_+^2)\nonumber\\
        &+16\gamma_-^2\gamma_+^2\om^4\left(G^{0}(\om,n)^2-G^{1}(\om,n)^2\right)^2,
\end{align}
where
\begin{equation}
  \label{023006-23}
  G^s(\om,n)=G_{0,sn}(\om,n), \quad s=0,1,
  \end{equation}
and
 \begin{equation}
   \label{020107-23}
   G_{x,y}(\om,n)=\frac{1}{n+1}\cdot\frac{1}{\om_0^2-\om^2}
   +\frac{2}{n+1}\sum_{j=1}^n\frac{
\cos(\frac{\pi   j(2x+1)}{2(n+1)}) \cos(\frac{\pi   j(2y+1)}{2(n+1)})}{\om_j^2-\om^2},\quad x,y\in\bbI_n
   \end{equation}
is the Green's functions of $-\Delta_N+\om_0^2-\om^2$,
 {and $\pm\om_j$, $j=0,\ldots,n$ are 
  the eigenvalues  of $-\Delta_N+\om_0^2$ defined by $\omega_j = \omega\left(\frac{j}{n+1}\right)$
  where $\omega(r)$ is given by \eqref{032806-23}.}

 {  It is easy to see from \eqref{020206-23} that $4\om^2\ga_-N\le
  D+\ga_-^2G^{1}(\om,n)^2$. Therefore, the following bound can be found
 \begin{equation}
  \label{011307-23}
W(\om,n)\le\frac{\big(\om
  F\big)^2}{4}\Big(\frac{1}{\ga_-}+\frac{1}{\ga_+}\Big),\quad n=1,2,\ldots.
\end{equation}
}

The  functions $G^s(\om,n)$  can be computed explicitly:
\begin{equation}
  \label{Green-s}
G^s(\om,n)=\frac{1}{n+1}\cdot\frac{1}{\om_0^2-\om^2}+\frac{2}{n+1}\sum_{j=1}^n\frac{(-1)^{j s}\cos^2(\frac{\pi j}{2(n+1)})}{\om_j^2-\om^2}\quad,\quad s=0,1,
\end{equation}



There are very different behaviors of $W(\om,n)$ depending on whether $\om$ is in
the spectrum of the harmonic chain, or not, see Figure  \ref{wwork1}.


\begin{center}
\includegraphics[width=7cm]{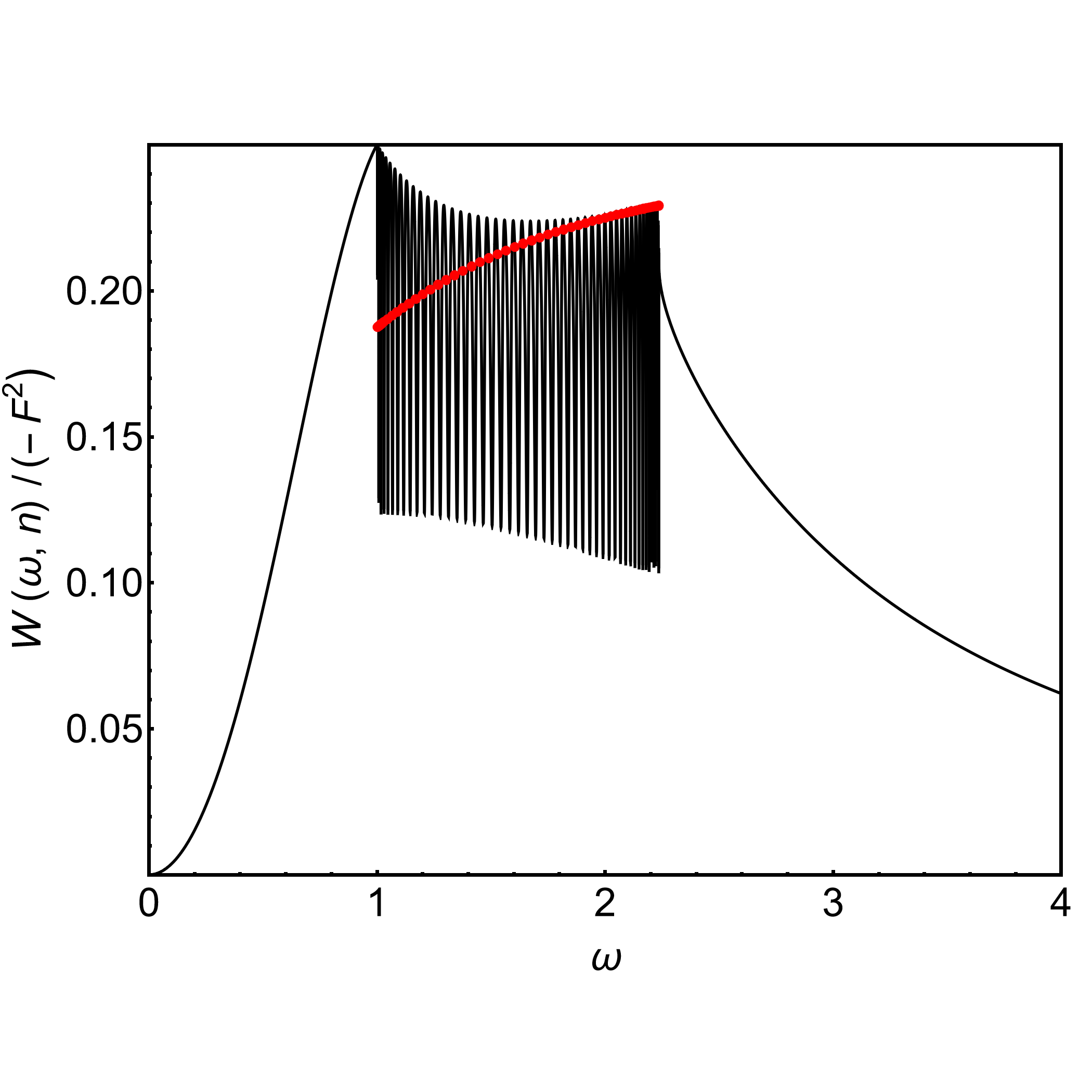}
\includegraphics[width=7cm]{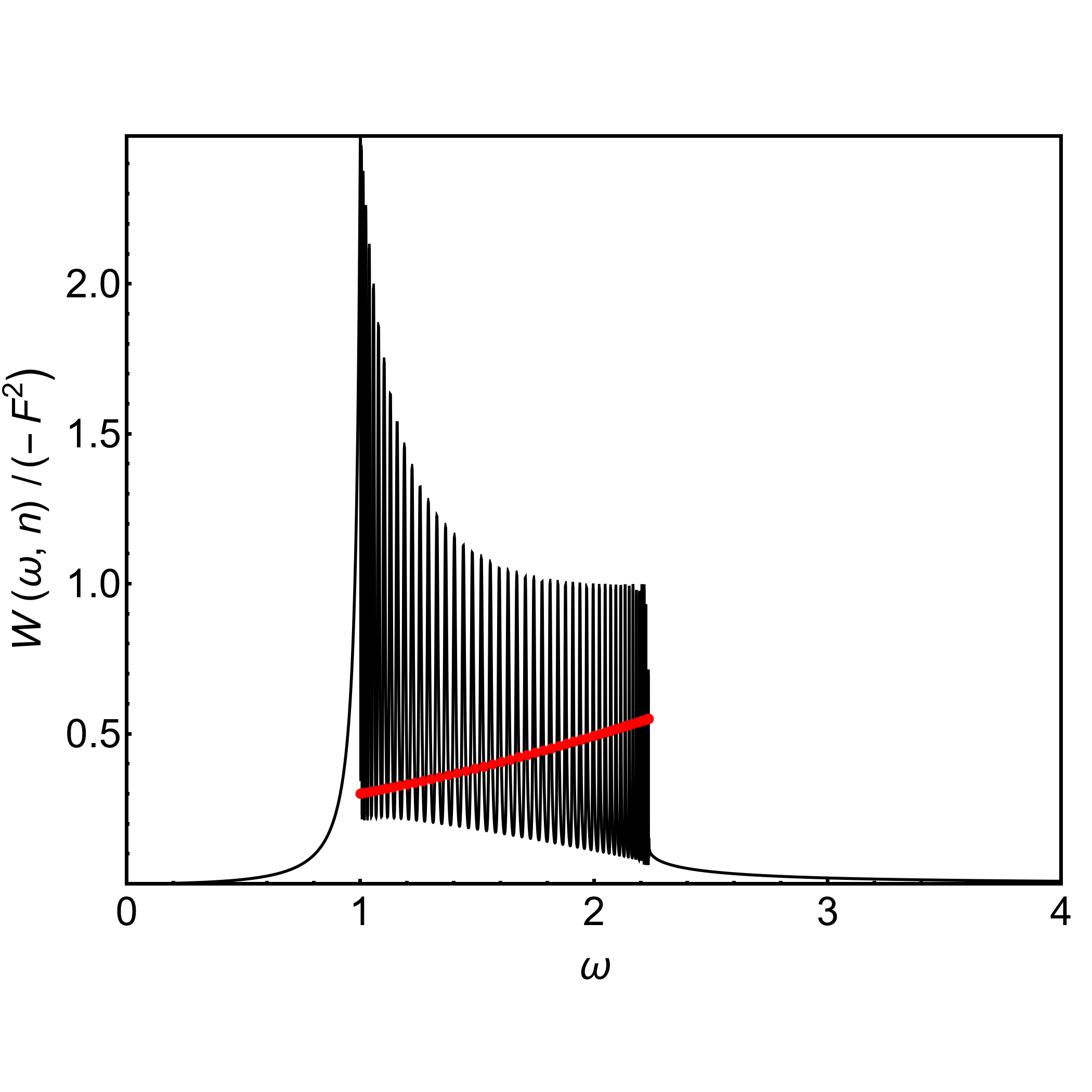} 
\vskip -0.5cm
\captionof{figure} {Behavior of the work for $\om_0=1$, $n=50$ with
  $\gamma_\pm=1$ (left figure) and $\gamma_-=1$, $\gamma_+=1/10$
  (right figure).  {The red points are the values of work computed at the points $\omega_j$
  of the harmonic spectra
  using equation \eqref{he}}. 
   Note
  the larger magnitude of the vertical scale on the right diagram.\label{wwork1}}         
\end{center}

In particular, the formula \eqref{033105-23} cannot be applied directly  when
  $\om=\om_j$ as then  both $G^s(\om,n)$, $s=0,1$ are   divergent. 
However, we can still use  the formula  to find $W(\om_j,n)$, because
both $N(\om,n)$ and $D(\om,n)$ have the same order of magnitude  in the
neighborhood of $\om_j$ and, due to the cancellation, the work remains finite.

More precisely, assume that given  $j $ we have $\om^2=\om^2_{j}+\epsilon$
for some $\eps\ll1/n$.  The Green's functions  can be then written in
the form
\begin{equation}
G^s(\om ,n)=\frac{2 (-1)^{js}}{n+1}\cos^2\left(\frac{\pi j}{2(n+1)}\right)\epsilon^{-1}+\bar G^s(\om ,n)
\end{equation}
where $\bar G^s(\om,n)$  is  of order $O(1)$ for $\eps\ll1/n$.
We obtain then
\begin{equation}
W(\om_j,n)=\frac{F^2}{4}\cdot\frac{\gamma_++\gamma_-+16\gamma_+\gamma_-^2\om_j^2 S(\om_j,n)}{(\gamma_++\gamma_-)^2+16\gamma_+^2\gamma_-^2\om_j^2 S(\om_j,n)}\label{he}
\end{equation}
where
\begin{equation}
S(\om_j,n)=\bar G^0(\om_j,n)-(-1)^j\bar G^1(\om_j,n).
\end{equation}

In particular, \eqref{he} implies that
$$
W(\om_j,n)\sim \frac{F^2}{4\gamma_{-}, },\quad\mbox{ as } \gamma_{+}\rightarrow
0\quad\mbox{ and }\quad 
W(\om_j,n)\sim \frac{F^2}{4\gamma_{+}, },\quad\mbox{ as } \gamma_{-}\rightarrow
0.
$$

 {
If $\gamma_+ = 0$ the formula \eqref{033105-23} for the work simplifies to
\begin{equation}
  \label{eq:3}
  W(\omega,n) = \left(\omega F\right)^2
  \frac{\gamma_- G^1(\omega,n)^2}{1+4\gamma_-^2\omega^2 G^0(\omega,n)^2}
\end{equation}
that gives $ W(\omega,n) \to 0$, as $\gamma_- \to 0$, if $\omega \neq \omega_j$.
This means that outside the resonance frequences, no work is done on the system
if dissipation is absent. Recall also that when $\gamma_+ = \gamma_-= 0$
{\sout{and $\omega - \omega_j$}}
the stationary periodic state does not exist as the energy keeps accumulating inside the system.

}

\subsubsection{Work in the case $n\to+\infty$ when $\om$  lies outside
  the harmonic
  chain spectrum}

Consider now the case $n\gg1$. The spectrum
{becomes then the interval}
$
{\cal I}:=[\om_0,\sqrt{\om_0^2+4}].
$ For $\om$ outside ${\cal I}$
the right hand side of the  formula for the Green's function, see
\eqref{Green-s}, does not contain any singular term and  
$G^s(\om,n)$   can be approximated by:
\begin{equation}
  \begin{split}
    G^0(\om,n) &= { 2\int_0^1 \frac{\cos^2\left(\frac{\pi r}{2}\right)}{\omega_0^2 -\omega^2
      + 4 \sin^2\left(\frac{\pi r}{2}\right)} \; dr+ O\Big(\frac{1}{n}\Big)}\\
    &=-\frac{1}{2}+\frac{1}{2\pi}\left(\om_0^2+4-\om^2\right)\int_0^\pi
    \,\frac{\dd q}{\om_0^2+2-\om^2-2\cos q}+O\Big(\frac{1}{n}\Big).
  \end{split}
\end{equation}

Using \cite[formula 2.553.3]{GR} we get \footnote{Note that formula
  \eqref{010809-23} makes also sense in case $\om_0=0$, as then 
  any $\om$ outside ${\cal I}$ satisfies $\om^2>4$.}
\begin{equation}
  \label{010809-23}
\bar
G^0(\om)=\lim_{n\rightarrow\infty}G^0(\om,n)=-\frac{1}{2}+\frac{\vert
  \om_0^2+4-\om^2\vert}{2\sqrt{(\om_0^2-\om^2)(\om_0^2+4-\om^2)}}\quad
\om\notin {\cal I}.
\end{equation}

Likewise, we can show
\begin{equation}
\bar G^1(\om)=\lim_{n\rightarrow\infty}G^1(\om,n)=0, \quad \om\notin
{\cal I}.
\end{equation}
Combining the above the work corresponding to $\om$
outside  the harmonic spectra is given by
\begin{equation}\label{eq:awork}
\bar W(\om)=\lim_{n\rightarrow\infty}W(\om,n)=\frac{\gamma_+\Big(FH\Big)^2 \Big[1+4\Big(\gamma_-H\Big)^2\Big]}{1+4(\gamma_+^2+\gamma_-^2)H^2+16\Big(\gamma_+\gamma_- H^2\Big)^2},
\end{equation}
where
\begin{align}
   \label{H}
  H(\om):=\om\bar G^0(\om).
 \end{align}
Observe that $\bar W(\om)$
tends to $0$, when $\ga_+\to0$.
 {
Likewise  $\bar W(\om) \to 0$, when either $\omega \to\infty$ or $\omega\to 0$.
Notice that there is still a strictly positive work done even if $\omega \notin \cal I$,
  as long as there is dissipation on the point where work is applied ($\gamma_+ >0$) and $\omega$
  is finite. We will see in \autoref{sec:curr-mech-energy} that this work flows directly into the right reservoir
  while the current of mechanical energy through the system vanishes as $n\to\infty$.
  In particular, it follows from \eqref{eq:awork} that}
 {  \begin{equation}
    \label{eq:12}
    \begin{split}
    \lim_{\om \uparrow \om_0} \bar W(\om) &=\bar W(\om_0):= \frac{F^2}{4 {\gamma_+}},
    \\
    \lim_{\om \downarrow \sqrt{\om_0^2 +4}} \bar W(\om) &= \bar
    W(\sqrt{\om_0^2 +4})\\
    &=\frac{\gamma_+ F^2 (\omega_0^2 +4) }{4}\cdot
    \frac{1+ \gamma_-^2 (\omega_0^2 +4)}
    {1+ (\gamma_-^2 + \gamma_+^2) (\omega_0^2 +4) +
      \gamma_-^2  \gamma_+^2 (\omega_0^2 +4)^2}.
  \end{split}
\end{equation}
This helps to understand the different scales {on vertical lines} in Figure \ref{wwork1}
 depending on the value of $\gamma_-$.}

 \subsubsection{The case $n\to+\infty$ and $\om$  is inside of the harmonic
  chain spectrum}

The computation of the $n\rightarrow\infty$ limit for the Green's
functions when $\om$ is inside the harmonic spectral interval ${\cal
  I}$ is more
complicated because there are singularities at the harmonic
frequencies $\om_j$ and the
distance between singularities is of order $1/n$.


Fix $\om$ inside  of ${\cal I}$.  To  describe the behavior of
    $W(\om,n)$  near the selected frequency $\om$ we  introduce
a function $\bar W(r,u)$, see  formula \eqref{033105-23a}.   This
function   is $1$-periodic in both variables and satisfies
$W(\om,n)= \bar W\big(r(\om),(n+1)r(\om))+o(1)$,
as $n\to+\infty$.
The description of $W(\om,n)$  in terms
 of the associated family of Young measures is given in \eqref{young}
 below.
 The work $W(\om,n)$ in the limit, when
 $n$ is large, is plotted in Figure \ref{wo1}.
\begin{center}
\includegraphics[width=6cm]{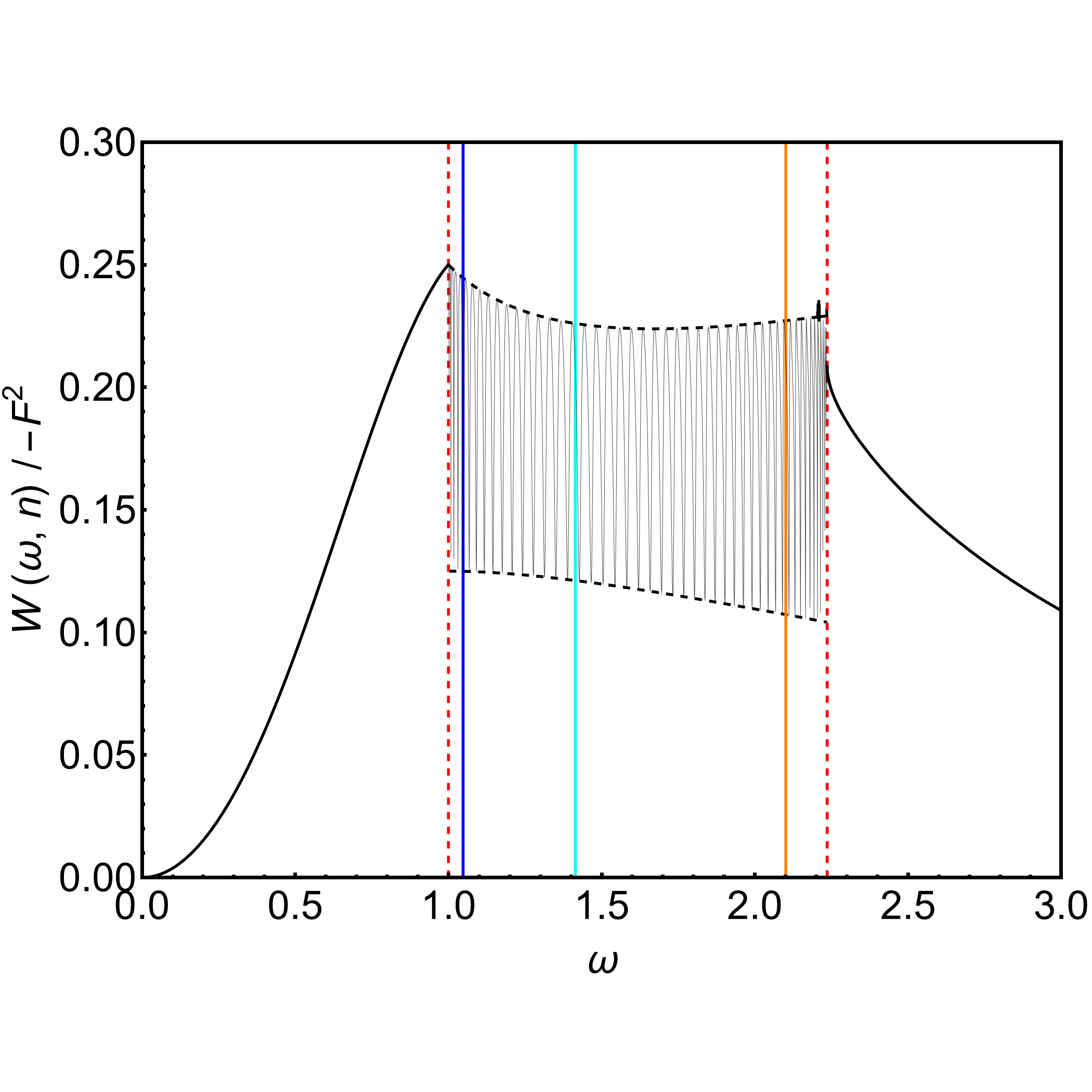}  
\includegraphics[width=6cm]{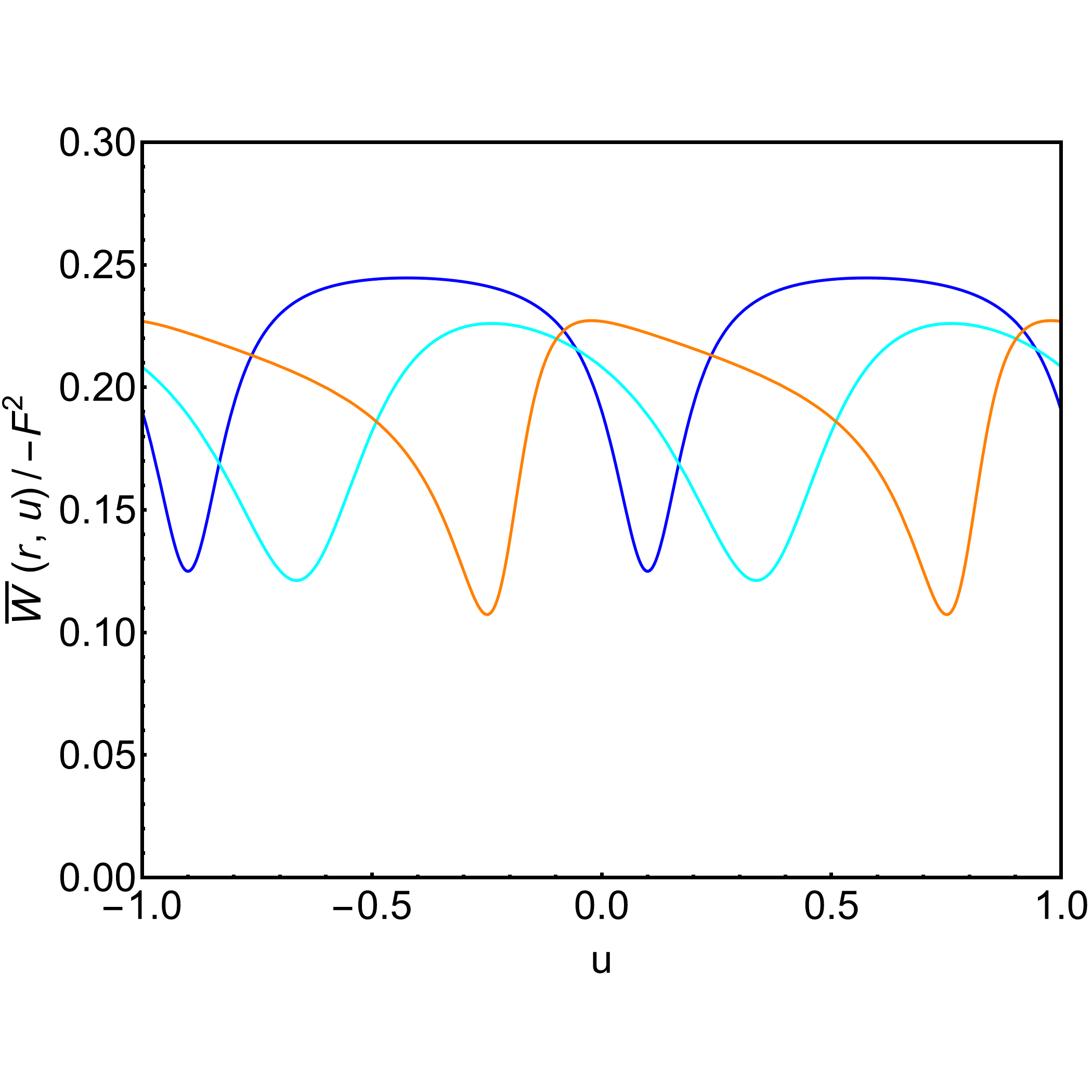}
\includegraphics[width=6cm]{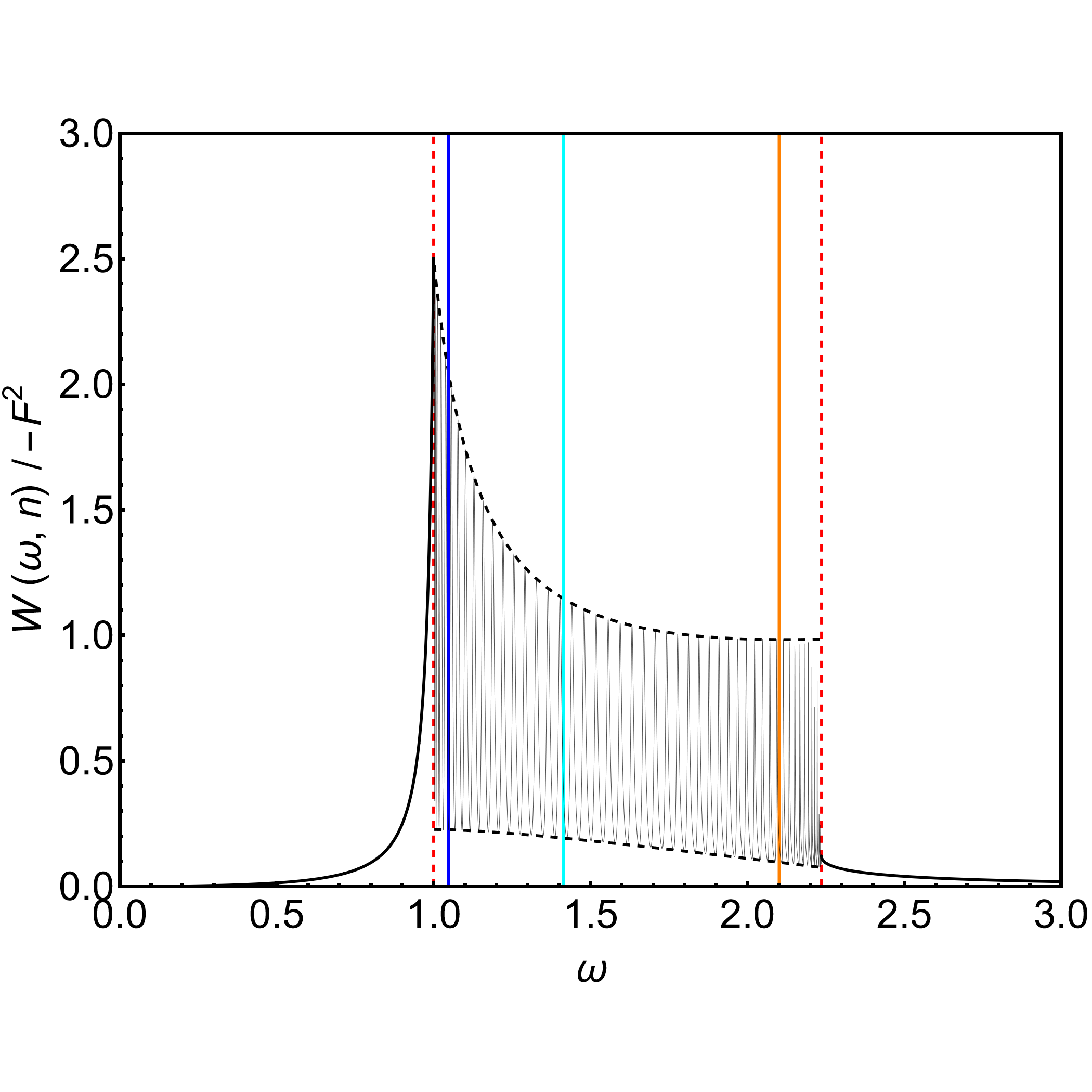} 
\includegraphics[width=6cm]{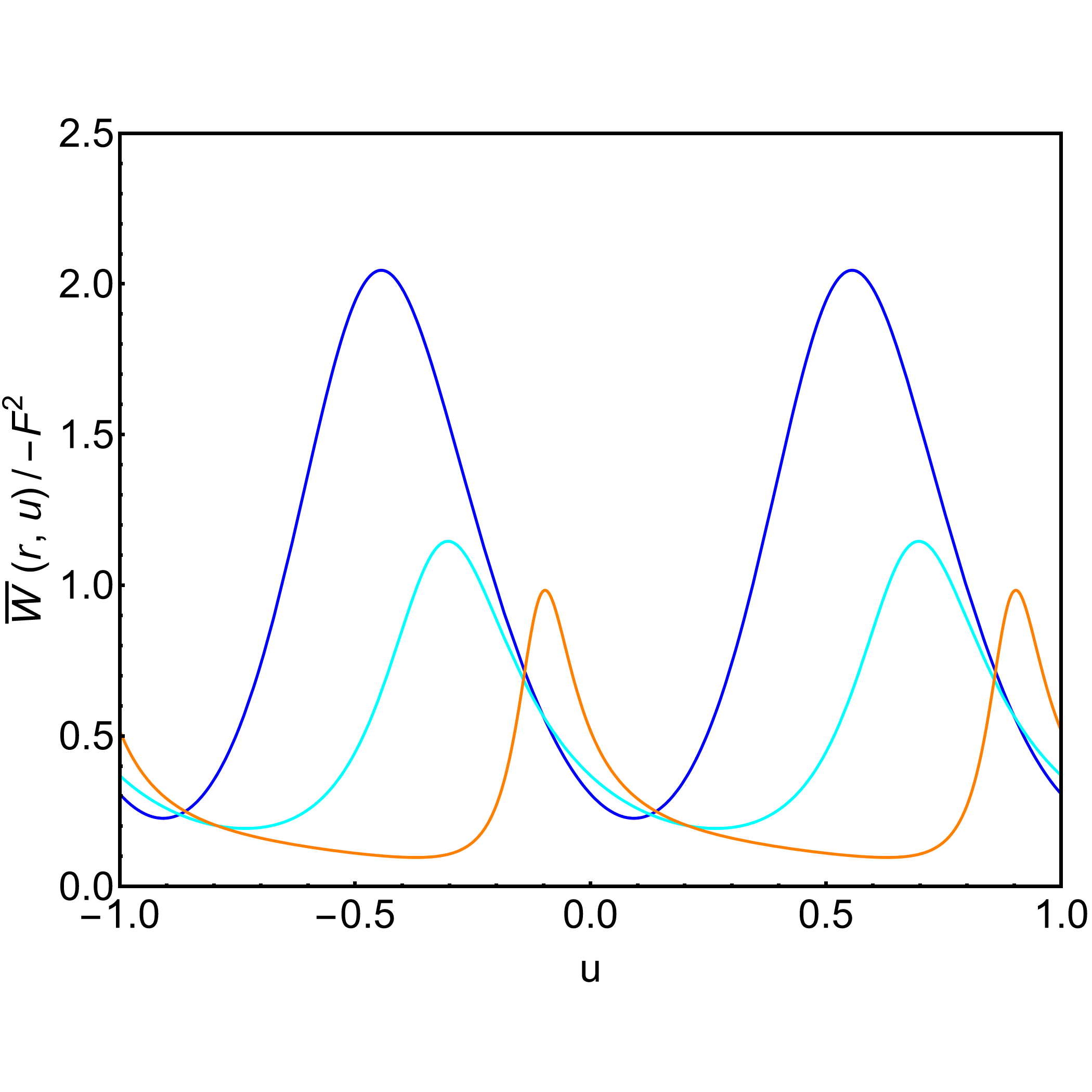}
\captionof{figure} {\small Behavior of the work functional. First row:
  $(\gamma_-,\gamma_+)=(1,1)$.
  Second row: $(\gamma_-,\gamma_+)=(1,1/10)$.
  Left column: work inside the harmonic spectrum   computed using  limiting expression
  \eqref{033105-23}  for
  $n\rightarrow\infty$. Black dotted curve represents $W(\om,n)$
    with $n=50$. Red dashed lines  stand for the limit of the harmonic
    spectrum. Blue, cyan and orange lines indicate the harmonic
    frequencies $\om=1.0478$, $1.41421$ and $2.101$,
    respectively. Right column: diagrams of  $\bar W\big(r,u\big)$,
    $u\in[-1,1]$  around the harmonic frequencies $\om=1.0478$ ({$r=0.1$},
    blue), $1.41421$ ({$r=0.66$}, cyan) and $2.101$ ({$r=0.75$}, orange). Note
  the larger magnitude of the vertical scale in the second row.\label{wo1}}         
\end{center}



 \subsubsection{The case of a general periodic force}

 {Finally, we remark that in the general case of a $\theta$-periodic force of the form
\begin{equation}
  \label{022905-23l}
\cF(t/\theta)=\sum_{\ell=1}^{+\infty}F_{\ell}\cos(\om(\ell) t),\quad
\mbox{where }\om(\ell):=\frac{2\pi\ell }{\theta}
\end{equation}
 whose real valued Fourier coefficients satisfy
$\sum_{\ell=1}^{+\infty}(\ell F_{\ell})^2<+\infty$,  the work
performed  by the
force     can be determined from the formula:
\begin{equation}
  \label{033105-23p}
W(n)=\sum_{\ell=1}^{+\infty}\big(\om(\ell) F_{\ell}\big)^2 \frac{N(\om(\ell),n)}{D(\om(\ell),n)}.
\end{equation}
Therefore its behavior, as $n$ gets large, can be determined from the term by term
analysis of   the series appearing on the right hand side
of \eqref{033105-23p}.
}



\subsection{Energy}

As in Section \ref{work} we assume that the periodic force ${\cal
  F}(t)$ is given by   \eqref{022905-23}.  The time average
of the expectation of the total energy energy of the chain  $E(\om,n)$
breaks up into the sum of thermal component $E_{\rm th}(\om,n)=\sum_{x\in\bbI_n}\lang e_x^{\rm th}\rang$
and the mechanical one $E_{\rm mech}(\om,n)=\sum_{x\in\bbI_n}\lang
e_x^{\rm mech}\rang$, with $  e_x^{\rm th} $ and $  e_x^{\rm mech} $
defined in \eqref{e-thm1} and \eqref{e-kin}, respectively.

Considering the behavior of  the thermal   energy functional, defined
in \eqref{e-thm}, it has been shown in
\cite{RLL67}, that in the case $\om_0=0$ and $\ga_-=\ga_+$ we have
$\lang e_x^{\rm th}\rang=\frac12(T_-+T_+)$ for all $x=1,\ldots,n-1$. 
If $\om_0>0$ and $\ga_-=\ga_+$, then 
\cite[formulas (38) and (42)]{Nak70} give
$$\lang e_x^{\rm
  th}\rang=\frac12(T_-+T_+)(1+o_x), \quad\mbox{where }|o_x|\le
\frac{C}{g^{x\wedge(n+1-x)}}
$$
for some constants $C>0$, $g>1$ independent of $n$. As a result we
have $E^{\rm th}(\om,n)\sim n$, as $n\to+\infty$.

 \subsubsection{Formula for the total mechanical energy functional for a single mode oscillating force}

In what follows we consider  the behavior of the mechanical component  of
the energy. Again, assume that   the force is given by
\eqref{022905-23}. It turns out, see Section \ref{appb} of the
Appendix,
that the  time average over the
period of the microscopic   mechanical energy density equals
\begin{equation}
  \label{013006-23}
 \lang e_x^{\rm mech}\rang
 =\frac{F^2}{2}\cdot\frac{M_x(\om,n)}{D(\om,n)},
 \end{equation}
 where $D(\om,n)$ is given by \eqref{020206-23} and
 $$
 M_x(\om,n)=G^1_x(\om,n)^2(\om^2+\om_0^2)+(\nabla^\star
 G^1_x)(\om,n)^2+(2\om\ga_-)^2 \Big[ {\cal G}_x(\om,n)^2 +(\nabla^\star {\cal G}_x)(\om,n)^2\Big],
 $$
 with (see \eqref{020107-23})
 \begin{equation}
   \label{030107-23}
   G^0_x(\om,n)=G_{0,x}(\om,n)\quad\mbox{ and
   }\quad G^1_x(\om,n)=G_{x,n}(\om,n).
 \end{equation}
 Using \eqref{020107-23} we get
 \begin{align*}
   G^s_x(\om,n)=\frac{1}{n+1}\cdot\frac{1}{\om_0^2-\om^2}
     +\frac{2}{n+1}\sum_{j=1}^n\frac{(-1)^{j s}\cos(\frac{\pi j}{2(n+1)})\cos(\frac{\pi j(2x+1)}{2(n+1)})}{\om_j^2-\om^2}\quad,\quad s=0,1
\end{align*}
and (recall $G^s=G^s_0$, $s=0,1$) 
$$
{\cal G}_x(\om,n)=G^0(\om,n) G^1_x(\om,n)-G^1(\om,n) G^0_x(\om,n).
$$
The explicit formula for the 
total mechanical energy functional, obtained by summing over all $ x$
expression \eqref{013006-23},
is presented in  \eqref{etot} below.

\subsubsection{Energy in the case $\om$  lies outside harmonic
  chain spectrum}

Analogously as in the case of the work functional the behavior
$E_{\rm mech}(\om,n)$ depends on whether the force frequency belongs to
the inside   or outside of 
 the spectrum of the harmonic chain. If $\om \not \in{\cal I}$ the
 asymptotics of $E_{\rm mech}(\om,n)$, as $n\to+\infty$,  can be obtained
 by a Riemann sum approximation. Then,
 \begin{equation}
  \label{040201-23b}
 \lim_{n\to+\infty}    E(\om,n)=\frac{F^2\left\{1
  +4\big(\ga_-H \big)^2\right\}}{4[1+ 4(
   \ga_-^2 +\ga_+ ^2)  H^2
    +16(\ga_+\ga_-   H^2)^2] }
 \Big[K_0 \left(\om^2+\om_0^2\right)+K_1\Big].
\end{equation}
  Here  $H$ is given by \eqref{H} and
\begin{align*}
   K_0=  \frac{\dd H}{\dd \om^2},\qquad
   K_1= \frac{\dd }{\dd \om^2}\left(\Gamma_0(\om)-\Gamma_2(\om)\right),
\end{align*}
where $\Gamma_x(\om)$ is the Green's function of the lattice $\bbZ$
laplacian. It is given by 

\begin{align*}
  &\Gamma_x(\om)=
   \left\{\Big[\om_0^2 -\om^2\Big]\Big[ 4+ \om_0^2 -\om^2\Big]\right\}^{-1/2}\\
  &
    \times \left\{1+\frac12\Big[\om_0^2
    -\om^2\Big]+\frac12\left\{\Big[\om_0^2
    -\om^2\Big]\Big[4+\om_0^2
    -\om^2\Big]\right\}^{1/2}\right\}^{-|x|},\quad \mbox{when $\om_0^2 >\om^2$ }
\end{align*}
and
\begin{align*}
  &\Gamma_x(\om)=-
   \left\{\Big[\om^2-\om_0^2 \Big]\Big[\om^2-\om_0^2-4 \Big]\right\}^{-1/2}\\
  &
    \times \left\{1-\frac12\Big[\om^2-\om_0^2 \Big]-
    \frac12
  \left\{\Big[\om^2-\om_0^2 \Big]\Big[\om^2-\om_0^2-4
  \Big]\right\}^{1/2} \right\}^{-|x|},\quad \mbox{when $\om_0^2 +4 <\om^2$.}
\end{align*}
Note that when $\ga_+\to0$, the formula \eqref{040201-23b} simplifies
and we have
\begin{equation}
  \label{040201-23bb}
 \lim_{n\to+\infty}    E(\om,n)=\frac{F^2 }{4 }
 \Big[K_0 \left(\om^2+\om_0^2\right)+K_1\Big].
\end{equation}

\subsubsection{The case when  $\om$ is inside of the harmonic
  chain spectrum}

If,  $\om$   is inside of ${\cal I}$, the
time average of  $ E_{\rm mech}(\om,n)$ is  proportional
to the size of the system. After normalization  we obtain, see Section
\ref{appb} of the Appendix,
\begin{align*}
 \frac{1}{n}E(\om,n)=\bar E \big(r,(n+1)r\big)+o(1)
     \end{align*}
as $n\to+\infty$, where $\bar E \big(r,u\big)$ is  $1$-periodic in the
first and $2$-periodic in the second variable. It is described by  formulas \eqref{etot1} and
\eqref{etot2}. Here
$r$ is determined from $\om$ by formula  \eqref{032806-23}.

Behavior of the energy functional is illustrated in
Fig. \ref{energy1}.
\begin{center}
\includegraphics[width=5cm]{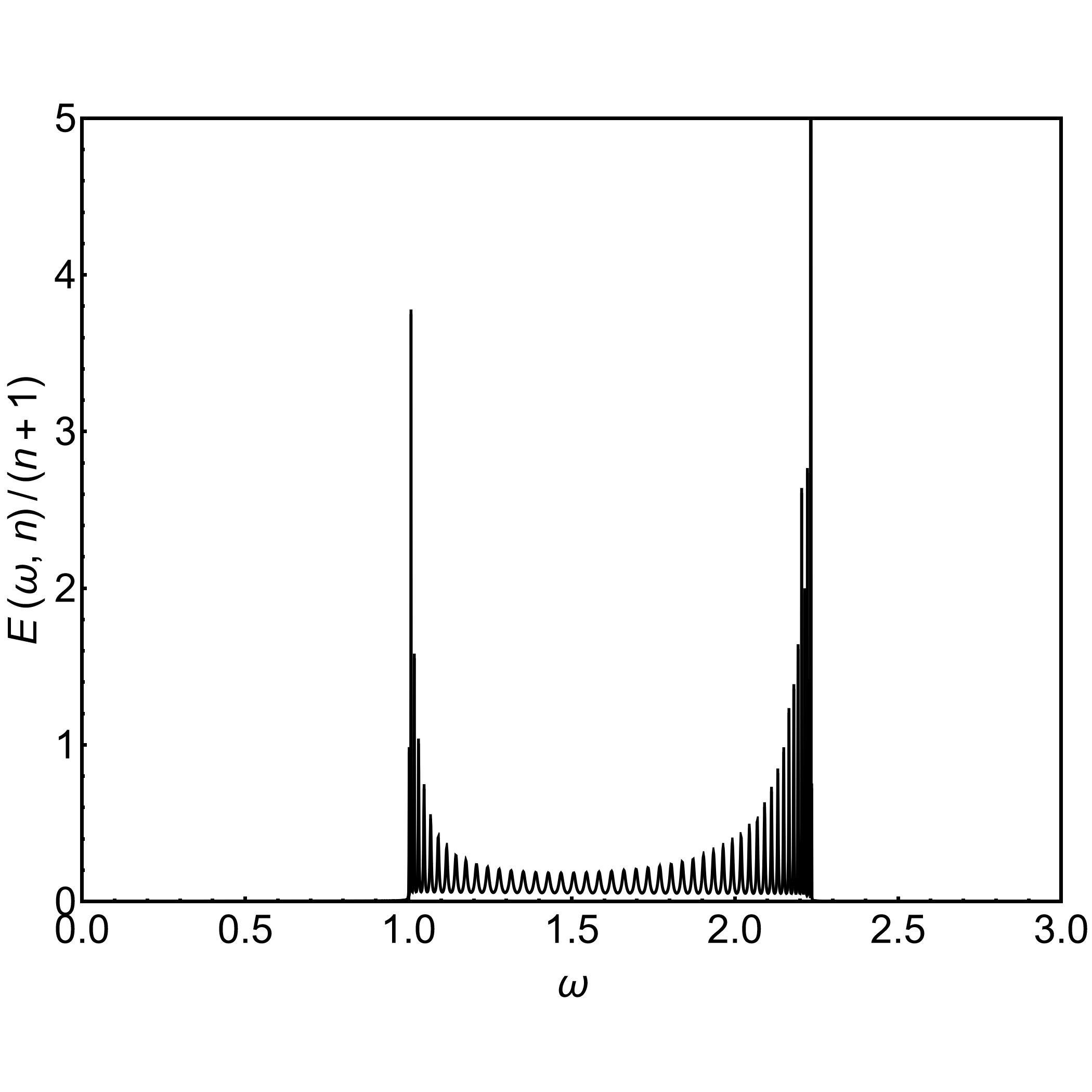} 
\includegraphics[width=5cm]{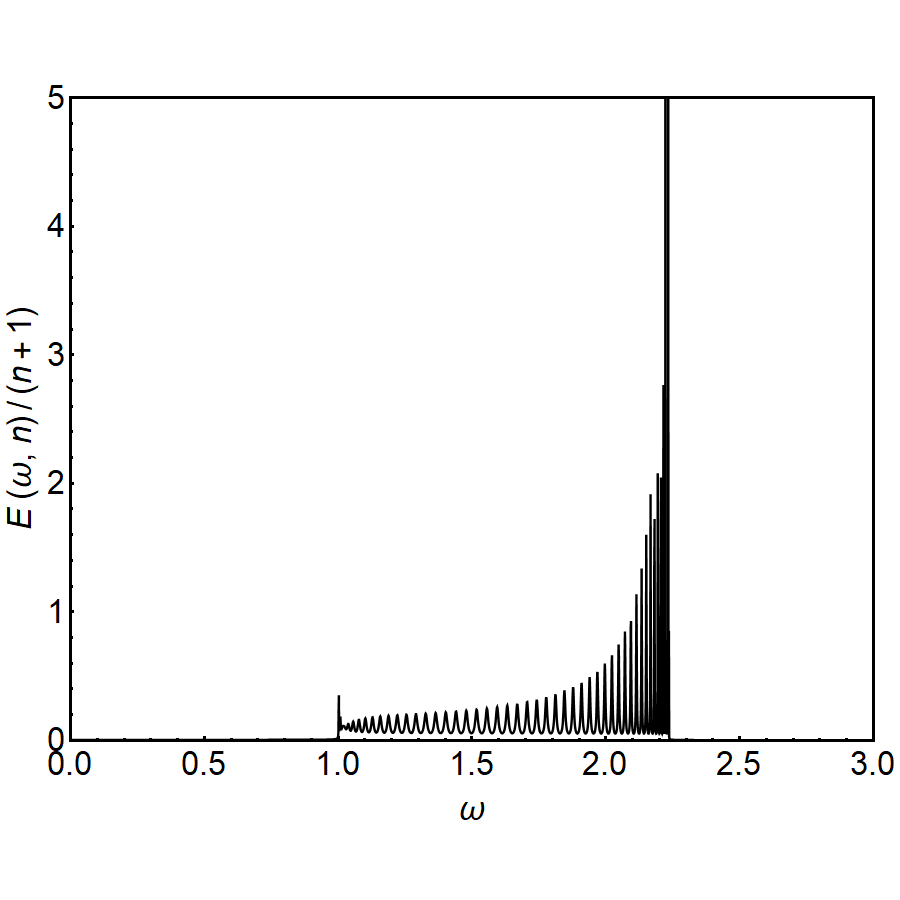}
\captionof{figure} {Behavior of the energy for $n=50$ with $\gamma_\pm=1$ (left) and $\gamma_+=1$, $\gamma_-=1/10$  (right). \label{energy1}}         
\end{center}

\section{Current of mechanical energy}
\label{sec:curr-mech-energy}

The currents of the mechanical energy are given by
\begin{equation}
  \label{eq:15}
  \begin{split}
    \bar{ j_{x,x+1}^{\text{mech}}}(t) &= - \bar p_x(t)\left(\bar q_{x+1}(t) - \bar q_{x}(t)\right) ,
    \qquad x=0,\dots,n-1\\
    \bar{ j_{-1,0}^{\text{mech}}}(t) &= -\gamma_- \bar p_0(t)^2, \qquad
     \bar{ j_{n,n+1}^{\text{mech}}}(t) = \gamma_+ \bar p_n(t)^2 - \mathcal F(t/\theta) \bar p_n(t).
  \end{split}
\end{equation}
 {They have all the same time average over the period:
\begin{equation}
  \label{eq:17}
  J^{\text{mech}}(n) := \lang\bar{ j_{x,x+1}^{\text{mech}}}\rang.
\end{equation}
Note that $W^-(n):=-J^{\text{mech}}(n)$ is the amount of work that
goes into the left reservoir. 
Of course when $\gamma_+ = 0$ we have $W^-(n)=  W(n)$.
If however $\gamma_+ >0$, then some of the work, denoted by $W^+(n)=
W(n)-W^-(n)$,  goes   into the right reservoir.

We compute first
$W^-(n)$, using $\bar{ j_{-1,0}^{\text{mech}}}(t)$,
as it involves   simpler formulas.
From \eqref{011012-22} we have
\begin{equation}
  \label{eq:18}
  \begin{split}
  &\bar p_0(t) = \text{Re}\left(i\om  {e^{i\om t}} \tilde q_0(\om)\right)\\
  &=  \frac{F G^1(\om,n) \left[\text{Re}(\tilde D(\om,n)) \om \cos(\om t)
   - \text{Im}(\tilde D(\om,n)) \om \sin(\om t)\right]}{\left| \tilde D(\om,n)\right|^2}
\end{split}
\end{equation}
and, recalling that $\om= \frac{2\pi}{\theta}$,
\begin{equation}
  \label{eq:19}
  \begin{split}
  J^{\text{mech}}(n) = -\frac{\gamma_-}{\theta} \int_0^\theta \bar
  p_0(t)^2 \dd t
    = - \gamma_-\Big( F\om G^1(\om,n)\Big)^2 
\end{split}
\end{equation}
As a result, combining with \eqref{033105-23}, we get
\begin{align}
  \label{011207-23}
    W^-(n)=\gamma_-\Big( F\om G^1(\om,n)\Big)^2,\qquad
    W^+(n)= W(n)- W^-(n).
  \end{align}
Notice that if $\om\notin \cal I$, since  $G^1(\om,n) \mathop{\longrightarrow}_{n\to\infty} 0$,
we have $J^{\text{mech}}(n) \mathop{\longrightarrow}_{n\to\infty} 0$.
Comparing with \eqref{eq:awork} we deduce that if $\om\notin \cal I$, all the work goes to
the right thermostat as $n\to\infty$.}

 {If $\om\in (\omega_0, \sqrt{\om_0^2+4})$ , then
$
    W^-(n)=\bar{W}\vphantom{1}^-\Big(r(\om),(n+1)r(\om)\Big)+o(1),
    $
    where the formula for $\bar{W}\vphantom{1}^- (r , u)$ can be obtained from
    \eqref{011207-23} by replacing $G^1(\om,n)$ by the function
    $\bar G^1(r,u)$ defined in \eqref{gr1}. We also have $
    W^+(n)=\bar{W}\vphantom{1}^+\Big(r(\om),(n+1)r(\om)\Big)+o(1),
    $ where $\bar{W}\vphantom{1}^+ (r , u) =\bar{W}(r , u)-\bar{W}\vphantom{1}^- (r , u)$, where
  $\bar{W}(r , u)$ is given by \eqref{033105-23a}.}

\appendix

\section{Time harmonics of the position and momenta averages}

\label{secc}

Recall that ${\cal F}(t/\theta)={\rm Re}\Big(Fe^{i\om t}\Big)$.
Consider the Fourier coefficients of  the means of the positions and momenta 
\begin{equation}
  \label{eq:6a}
\tilde p_x =\frac{1}{\theta}\int_0^\theta e^{- i\om t}\bar
p_x(t)\dd t,\quad \tilde q_x=\frac{1}{\theta}\int_0^\theta e^{-
  i\om t}\bar q_x (t)\dd t,\quad x\in\bbI_n.
\end{equation}
We have
$\bar
p_x(t)={\rm Re}\Big(\tilde p_xe^{i\om t}\Big)$ and $\bar
q_x(t)={\rm Re}\Big(\tilde q_xe^{i\om t}\Big)$.

From \eqref{eq:flip} and \eqref{eq:rbdf} we obtain $\tilde p_x
=i\om \tilde q_x$ and 
\begin{equation} 
\label{eq:qdynamicsbulk-av-f}
  i\om\tilde   p_x  =\Big( \Delta_{\rm N}   -\om_0^2
  -2i\gamma_x \om\Big)\tilde   q_x+F\delta_{x,n}
    , \; \quad x\in \bbI_n.
\end{equation}
Here
$\gamma_x= \gamma_-\delta_{0,x}+
\gamma_+\delta_{n,x}$. Substituting into \eqref{eq:qdynamicsbulk-av-f}
for $\tilde p_x $ we get the equation 
\begin{equation} 
\label{011005-21}
  0 =  \Big(\Delta_x   
  +\om^2-\om_0^2  - 2i\om\gamma_x  \Big)\tilde q_x
  +  F\delta_{x,n}
    , \; \quad x\in\bbI_n.
  \end{equation}
 Hence, using the notation of \eqref{030107-23}, we can write
\begin{equation} 
\label{021205-21a}
 \Big(F-2i\ga_+ \om\tilde
    q_n \Big)G^1_x(\om,n) 
    -2i\ga_+ \om\tilde
         q_0 G^0_x(\om,n) =  \tilde
    q_x ,\quad x\in\bbI_n.
\end{equation}
For  $x=0,n$ we get a closed system of $2$ equations for $\tilde
         q_0$ and  $\tilde
         q_n$ that can be solved explicitly and we obtain
         \begin{equation} 
\label{011012-22}
  \tilde  q_0 =\frac{F G^1 (\om,n)}{\tilde D( \omega,n)},\quad \tilde
    q_n =\frac{F\tilde N(\pm\omega,n)}{\tilde D( \omega,n)}
   ,
\end{equation}
where, using the notation of \eqref{023006-23}, we have
\begin{align}
  \label{D}
&\tilde N(\omega,n)=G^0(\omega,n)+2i\omega\gamma_-\left(G^0(\omega,n)^2-G^1(\omega,n)^2\right)\nonumber\\
&\tilde
                                                                                                            D(\omega,n)=1-4\gamma_+\gamma_-\omega^2\left(G^0(\omega,n)^2-G^1(\omega,n)^2\right)+2i\omega(\gamma_++\gamma_-)G^0(\omega,n). 
\end{align} Substituting back into \eqref{021205-21a} we conclude that
\begin{equation} 
  \begin{split}
    \label{021205-21e}
& \tilde
    q_x    =F\Big(a G^0_x(\om,n) +b G^1_x(\om,n) \Big),\quad\mbox{where},\\
& a=1-2i\omega\gamma_+\frac{\tilde N(\omega,n)}{\tilde
                 D(\omega,n)},\quad
             b =-2i\omega\gamma_-\frac{G^1(\omega,n)}{\tilde D(\omega,n)} .
\end{split} 
\end{equation}
Using \eqref{013105-23} and the fact that
$\bar p_n(t)=-\om{\rm Im}\Big( \tilde
    q_ne^{i\om t} \Big)
    $
we  obtain  \eqref{033105-23}.

\section{Time average of work functional when $\om$ is inside ${\cal
    I}$ and  $n\to+\infty$}

\label{appa}

 {We consider now $\om\in (\om_0,\sqrt{\om_0^2+4})$.
  We will parametrize the spectrum using $r(\omega) \in (0,1)$, defined by \eqref{eq:11},
  and we study here the asymptotic behaviour of $\tilde W(r,n) =W(\om(r),n) $.
  Similarly we define $\tilde G^{s}(r,n), s=0,1$.

  Denote $j(r) = [(n+1) r]$ (where $[x]$ denotes the integer part of $x$) and
  \begin{equation}
    \label{eq:5}
    u(r) = (n+1) r - [(n+1) r] \in (0,1).
  \end{equation}
  Since we are choosing $\om \neq \omega_j$, we have that $ u(r)\in(0,1)$.
}.

To compute    {$\tilde G^0(r,n)$}
we start with extracting the singular term at $\om_j$. 
From \eqref{Green-s} we get  
\begin{equation}
  \label{G0}
 {\tilde G^0(r,n)} =\frac{1}{n+1}\cdot\frac{1}{\om_0^2-\om(r)^2} +I_-(r;0,j-1)+I_+(r;1,n-j),
\end{equation}
where 
\begin{equation}
  \label{012806-23}
  I_\pm(r;m,k)=\frac{2}{n+1}\sum_{\ell=m}^{k}
  \frac{\cos^2(\frac{\pi (j(r) \pm \ell)}{2(n+1)})}{\om\left(\frac{j (r)\pm \ell}{n+1}\right)^2 -\om(r)^2},
\end{equation}
For any $1\le k_0\le k$ we break $I_\pm(\om;1,k)$ in two terms: one with
the first $k_0$ terms and the other with the remaining $k-k_0$
ones. The idea is to assume that $k$ is of order $n$ and $k_0$ is of
order $n^a$, with $a\in(0,1)$, when $n\rightarrow\infty$.  The first
term can be summed up explicitly and for the second we can use the
Riemann sum approximation, since  we are far away from the
singularity that occurs at $\om_j$. More precisely we can write
$
I_+(r;0,k)=I_+(r;0,k_0)+I_+(r;k_0+1,k)$.
Using the formula
\begin{equation}
  \label{cot}
  \cot(\pi x)=\frac{1}{\pi x}-\frac{2x}{\pi}\sum_{j=1}^{+\infty}\frac{1}{j^2-x^2},
\end{equation}
for $k_0 \sim n^a$, $a<1$, and large $n$ we have
\begin{equation}
  \begin{split}
    I_+(r;0,k_0)= \frac{1}{n+1}\sum_{\ell=0}^{k_0}
    \frac{\cos^2(\frac{\pi (j(r) + \ell)}{2(n+1)})}{\cos(\pi r) - \cos\left(\pi\frac{j(r) + \ell}{n+1}\right)}\\
    = \frac{1}{n+1}\sum_{\ell=0}^{k_0} \frac{\cos^2(\frac{\pi (j(r) + \ell)}{2(n+1)})}
    {-\sin\left(\frac{j(r) + \ell}{n+1}\right)\frac{\pi}{n+1}\left((n+1)r - j(r) - \ell\right)} + o(1)
    \\
     = - \frac{1}{2\pi}\sum_{\ell=0}^{k_0} \frac{\cot\left(\frac{\pi (j(r) + \ell)}{2(n+1)}\right)}
    {\left(u(r) - \ell\right)} + o(1)
   = \frac{\cot(\pi r/2)}{2\pi }\sum_{\ell=0}^{k_0}\frac{1}{\ell- u(r)} +  o(1).
\end{split}
\end{equation}
The sum in the last expression diverges, when $k_0\rightarrow\infty$. However, in the
expression \eqref{G0}  for $\tilde G^0$ we have also
\begin{equation}
I_-(r;1,k_0)=-\frac{\cot(\pi r/2)}{2\pi}\sum_{\ell=1}^{k_0}\frac{1}{\ell + u(r)}+o(1)
\end{equation}
and, as a result of the cancelation, the sum of
them has a finite limit as $k_0\rightarrow\infty$.  It can be computed
and the result is:
\begin{equation}
  \label{052806-23a}
\lim_{k_0\rightarrow\infty}\big(I_+(r;0,k_0)+I_-(r;1,k_0)\big)=
-\frac{1}{2}\cot(\pi r/2)\cot(\pi u(r)) .
\end{equation}
 
  Now we compute  the remaining expressions   $I_\pm(r;k_0+1,k)$  by
  using 
  the  Riemann sum approximation:
 \begin{equation}\label{eq:boh}
   \begin{split}
     &I_\pm( r;k_0+1,k)=
     \frac{2}{n+1}\sum_{\ell=k_0+1}^k
 \frac{\cos^2(\frac{\pi (j(r) \pm \ell)}{2(n+1)})}{\om\left(\frac{j(r) \pm \ell}{n+1}\right)^2-\om^2(r)}
    \\
 &=  2 \int_{v_0}^{v}
     \frac{\cos^2\left(\frac{\pi}{2}(r \pm \bar v)\right)}
     {\om(r \pm \bar v)^2- \om(r)^2}
     d\bar v
     +o(1)=   \frac12\int_{v_0}^{v}\,\frac{\big[1+\cos\big( \pi(r \pm \bar v)\big)\big]}
     {\cos(\pi r)-\cos(\pi(r \pm \bar v))}  dv  +o(1),
\end{split}
 \end{equation}
 where $v_0=k_0/(n+1)$, $v=k/(n+1)$. The last integral has a logarithmic
singularity when $v_0\rightarrow 0$ (i.e. $k_0\ll n$). Nevertheless, when putting
together the two terms, we obtain the principal value of the
integral at the singular point and, as a result,
{\begin{align}
      \label{012906-23}
& I(r)=\lim_{n\to+\infty}\Big(I_+(r;k_0,j) +I_-(r;k_0,n-j)\Big)   
     =
     \frac12 {\rm p.v.}\int_{-r}^{1-r}\,\frac{\big[1+\cos\big( \pi(r+\bar v)\big)\big]\dd\bar v}{\cos(\pi
    r)-\cos(\pi(r+
   \bar v))} \notag\\
  &
    =-\frac{1}{2}+ \cos^2\big( \pi r/2\big) {\rm p.v.}\int_{0}^{1}\,\frac{\dd\bar v}{\cos(\pi
    r)-\cos(\pi \bar v)}
    \end{align}
    (we recall that $k_0\simeq n^a$ with $a\in(0,1)$, and  {$j=[r (n+1)]$}).
Using \cite[formula  2.551.3, p. 171]{GR} 
$$
\int\,\frac{\dd\bar v}{\cos(\pi
    r)-\cos(\pi
   \bar v)}=\frac{1}{\pi\sin(\pi r)}\log\Big|\frac{(1+\sin(\pi
   r))/\cos(\pi r)-\tan\left(\frac{\pi}{2}\big(\bar v+\frac12\big)\right)}{(1-\sin(\pi
   r))/\cos(\pi r)-\tan\left(\frac{\pi}{2}\big(\bar v+\frac12\big)\right)} \Big|.
 $$
 we conclude that
 that the principal value of the integral on the utmost right hand
 side of \eqref{012906-23} equals null.
 Hence
 $I(r)
     = -1/2$.
}

Finally, putting together 
\eqref{052806-23a} and \eqref{012906-23}
we find:
\begin{align}
&  {\tilde G^0(r,n)=\bar G^0\big(r,  (n+1)r \big)} + o(1),\quad\mbox{where}\nonumber\\
&\bar G^0(r,u)=-\frac{1}{2}\Big(\cot(\pi r/2)\cot(\pi u) +1\Big).
   \label{010307-23}
\end{align}
We will consider $u\in \R$ and extend periodically the function $\bar G^0(r,u)$.

 

 We  compute $\bar G^1(r) =\lim_{n\to+\infty}\tilde G^1(r,n)$ by using formula:
\begin{equation}
  \frac{1}{2}\Big(\tilde G^0(r,n)+ \tilde G^1(r,n)\Big)=
  \frac{1}{n+1}\cdot\frac{1}{\om_0^2-\om(r)^2}+
  \frac{2}{n+1}\sum_{\ell=1}^{n/2}\frac{\cos^2(\frac{\pi \ell}{n+1})}{4\sin^2(\frac{\pi \ell}{n+1})
    +\om_0^2-\om(r)^2}.
\end{equation}
That is a very similar expression to the original one for
$\tilde G^0(r,n)$, see \eqref{Green-s} with $s=0$,
but with  factors $2$ not present in the denominators of fractions
appearing in the infinite sum.
{Following analogous arguments to the ones used before we find, in the limit $n\rightarrow+\infty$
  \begin{equation}
    \label{022906-23}
  \begin{split}
    &\frac{1}{2}\Big(\tilde G^0(r,n)+ \tilde G^1(r,n)\Big) =
    \bar H\Big(r, (n+1)r \Big) +o(1),\quad \mbox{where}\\
&  \bar H(r,u)= -\frac{1}{4}\Big(\cot(\pi r/2)\cot(\pi u/2) +1\Big).
    \end{split}
\end{equation}}
Therefore we get $\tilde G^1(r,n)=\bar G^1\big(r,  (n+1)r  \big)+o(1)$, as $n\to+\infty$,
where
\begin{equation}
  \bar G^1(r,u)=(-1)^{[u]+1}\frac{ \cot(\pi r/2)}{ 2\sin(\pi u)}.\label{gr1}
\end{equation}
We have shown therefore that
\begin{equation}
  \label{033105-23a}
  \begin{split}
    & \tilde W(r,n)= \bar W(r,(n+1)r) +o(1),\quad\mbox{where}\\
    &  {\bar W(r,u) =  F^2 \om(r)^2\frac{\bar N(r,u) }{\bar D(r,u) } }.
\end{split}
\end{equation}
The functions $\bar N(r,u)$
and $\bar D(r,u)$ are given by analogues of \eqref{020206-23}, with $G^s(\om,n)$
replaced by  $\bar G^s(r,u)$, respectively for $s=0,1$.  {As in
  \eqref{011307-23}  we get
  $$0\le \bar W(r,u)\le
  \frac{F^2}{4}\left(\frac{1}{\ga_-}+\frac{1}{\ga_+}\right).
$$}

 {Equality \eqref{033105-23a}  can be used to find
the family of Young measures  that is associated with
$\tilde W(r,n)$.
This is defined, see e.g. \cite[Section 1.E.3, p. 16]{evans}, as a family of measures $\mu(r,\dd v)$, $v\in\bbR$ such that
for any test function $\varphi\in L^1[0,1]$
and a bounded continuous function $\Phi\in C_b(\bbR)$
\begin{align}
  \label{young}
\lim_{n\to+\infty} \int_{0}^{1} \Phi\Big(\tilde W(r,n)\Big)
  \varphi(r)\dd r = \int_{0}^{1} \bar \Phi (r) \varphi(r)\dd r,
\end{align}
with
$
\bar \Phi (r) :=\int_\bbR  \Phi(v)\mu(r,\dd v).
$
Thanks to \eqref{033105-23a} we conclude that the probability measures
$\mu(r,\dd v)$
obtained by transporting the Lebesgue measure ${\rm m}(\dd u)$ on
$(0,1)$
by the mapping $u\mapsto \bar W\big(r,u\big) $ constitute the family of
Young measures associated with the sequence
 $\tilde W(r,n)$. We have
$
\mu(r,A)={\rm m}\Big[u: \bar W\big(r,u\big) \in A\Big]
$
for any Borel measurable subset $A$ of $\bbR$.
 {Since $\bar W\big(r,u\big)$ is bounded, piecewise  $C^1$-smooth and $\{u: \bar W\big(r,u\big) = v\}$}
 { is finite for each $r$,}  {the Young measures $\mu(r,\dd v)$,
   have  compactly supported
   densities.}
 {Using the frequency domain in the description of the Young
   measures, we conclude from \eqref{young} that
\begin{align}
  \label{young1}
\lim_{n\to+\infty} \int_{\cal I} \Phi\Big(  W(\om,n)\Big)
  \varphi(\om)\dd \om = \int_{\cal I} \bar \Phi (\om) \varphi(\om)\dd \om,
\end{align}
for any function $\varphi\in C_b({\cal I})$. Here $
\bar \Phi (\om) :=\int_\bbR  \Phi(v)\mu(r(\om),\dd v).
$ We have
\begin{align*}
 &\lim_{\om \downarrow \om_0 } \mu(r(\om),\dd v)=\delta_{\bar
   W(\om_0)}(\dd v),\\
  &\lim_{\om \uparrow \sqrt{\om_0^2+4} } \mu(r(\om),\dd v)= \delta_{\bar
    W(\sqrt{\om_0^2 +4}) }(\dd v),
\end{align*}
where, as we recall $\bar W(\om_0)$ and $\bar
    W(\sqrt{\om_0^2 +4})$ are given  in   \eqref{eq:12}.
The limit
holds in the sense of the weak convergence of measures.}
}

\section{Time average of energy in case $\mathbf \omega$  is  inside
  of ${\cal I}$}
\label{appb}

Formula \eqref{013006-23} is a direct consequence of  \eqref{e-kin} and formula \eqref{021205-21e}. 
Summing over all $x$ we conclude that
\begin{align}
  \label{etot}
E_{\rm mech}(\omega,n) &=F^2 \Biggl\{ (\vert a\vert^2+\vert b\vert^2)\biggl[
           (\omega^2+\omega_0^2)I_0(\omega,n)+J_0(\omega,n) \biggl]\\
&+{\rm Re}\big(a^*b) \biggl[
           (\omega^2+\omega_0^2)I_1(\omega,n)+J_1(\omega,n) \biggl] \Biggl\},
\end{align}
where  
\begin{align}
  \label{IJ}
I_s(\omega,n)=\frac{\dd}{\dd\omega^2}G_s(\omega,n),\quad
J_s(\omega,n)=\frac{\dd}{\dd\omega^2}R_s(\omega,n)
\end{align}
and
\begin{equation}
  \label{R}
R_s=\frac{2}{n+1}\sum_{j=1}^n\frac{(-1)^{j s}\sin^2(\frac{\pi j}{(n+1)})}{4\sin^2(\frac{\pi j}{2(n+1)})+\omega_0^2-\omega^2},\quad s=0,1.
\end{equation}
The remaining
terms have been defined in Section \ref{secc}.

The four functions: $I_s(\omega,n)$
and $J_s(\omega,n)$, $s=0,1$, appearing in  \eqref{IJ}
diverge, as 
$n\rightarrow\infty$, for $\om$ inside of ${\cal I}$.  Computations
involving  these functions use the same
technique as in the case of the asymptotics of the work
functional considered in Section \ref{appa} of the Appendix. We obtain
\begin{align}
  &  e(\omega,n) =\frac{E_{\rm mech}(\omega,n)}{n}= \bar e\big(r,(n+1)r\big)+o(1),\quad\mbox{where} \label{etot1}\\
  &
   \bar e\big(r,u\big), = F^2\left[(\omega^2+\omega_0^2)\left[(\vert \bar a\vert^2+\vert
    \bar b\vert^2)I_0\big(r,u\big)
+(\bar a^*\bar b+\bar a\bar b^*)I_1\big(r,u\big) \right]\right.\nonumber\\
&+\left.(\vert\bar  a\vert^2+\vert
                                                              \bar b\vert^2)J_0\big(u\big)+(\bar a^*\bar b+\bar a\bar b^*)J_1\big(u\big)\right]+o(1),
                                                              \notag
\end{align}
with the formulas for   terms $\bar a$ and $\bar b$ given by
analogues of \eqref{021205-21e}, where the Green's functions
$G^s(\om,n)$ are replaced by $\bar G^s(r,u)$, $s=0,1$, defined in
\eqref{010307-23} and \eqref{gr1}.
  Here $r$ is determined from $\om$ by eqt.
\eqref{032806-23} and 
\begin{align}
&I_s(r,u)=\frac{(-1)^{s[ u]}}{8}\cdot\frac{\cos(s\pi u)}{\sin(\pi r/2)\sin^2(\pi u)},\nonumber\\
&J_s(u)=\frac{ (-1)^{s[u]}}{2}\cdot\frac{\cos(s\pi u)}{\sin^2(\pi
            u)},\quad\quad s=0,1.
            \label{etot2}
\end{align}

Figure \ref{en7} illustrates the behaviour of $e(\omega)$.
\begin{center}
\includegraphics[width=6cm]{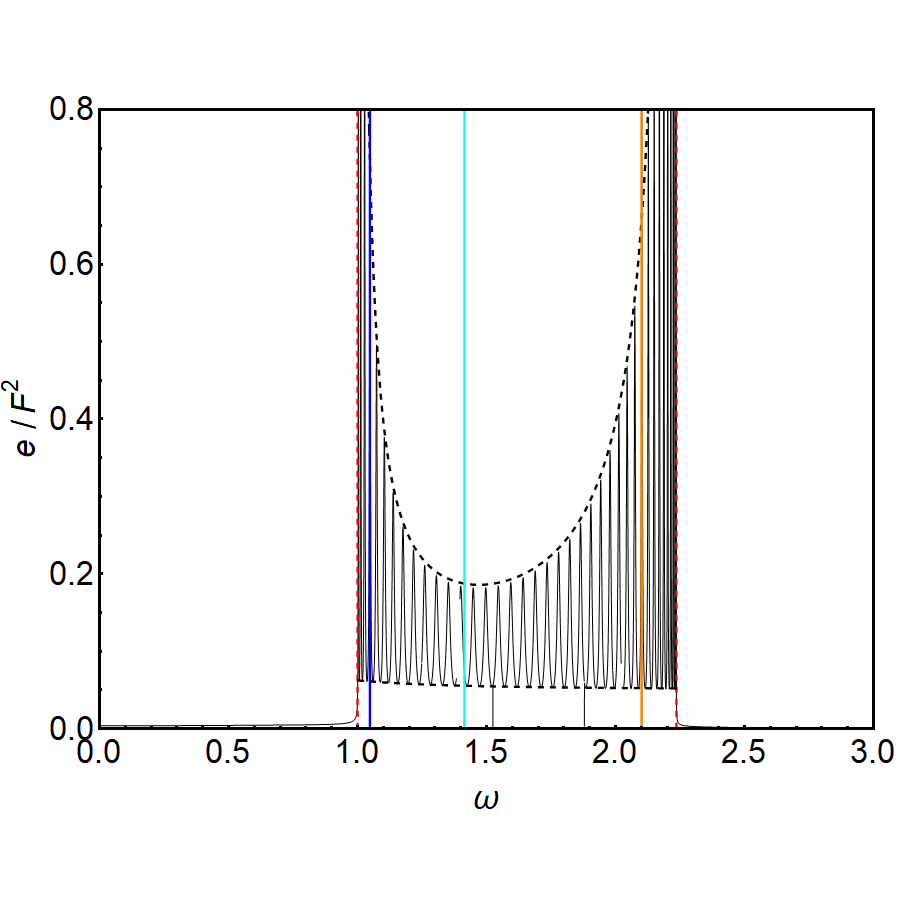}  
\includegraphics[width=6cm]{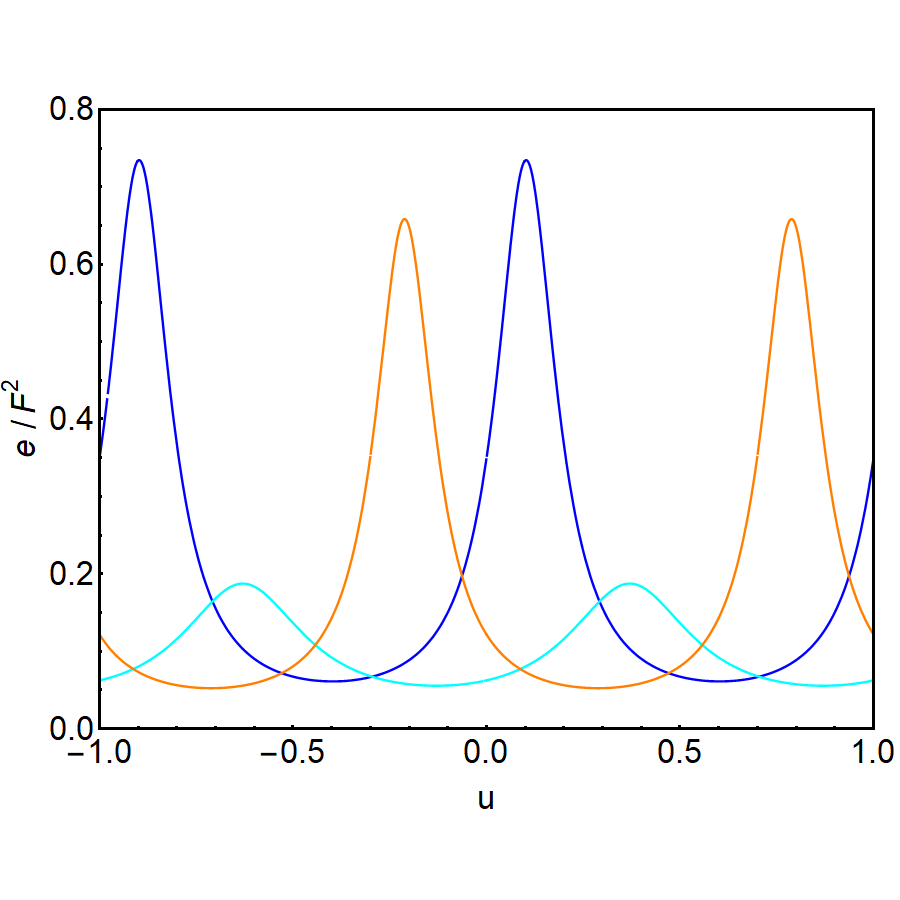}
\end{center}

\begin{center}
\includegraphics[width=6cm]{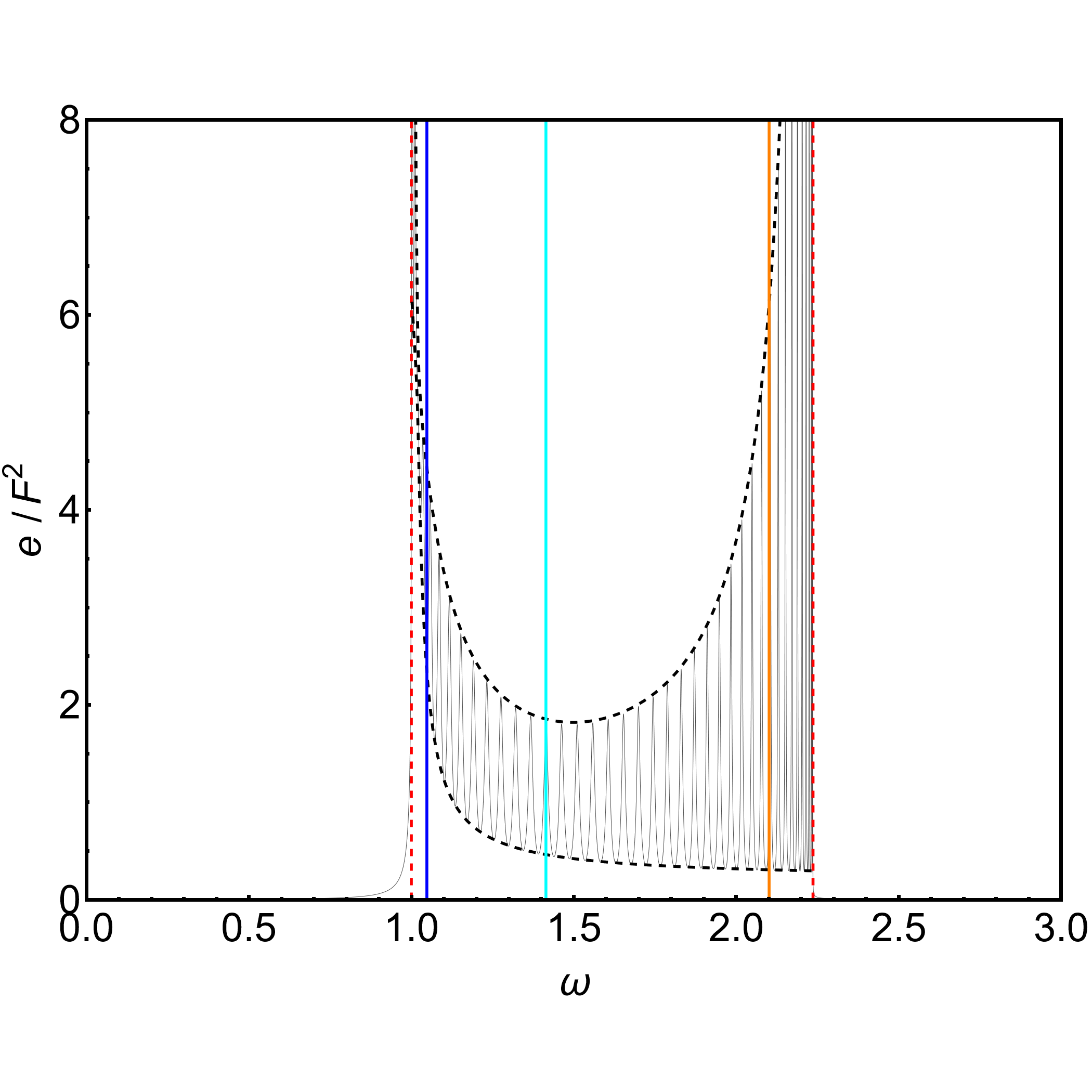}
\includegraphics[width=6cm]{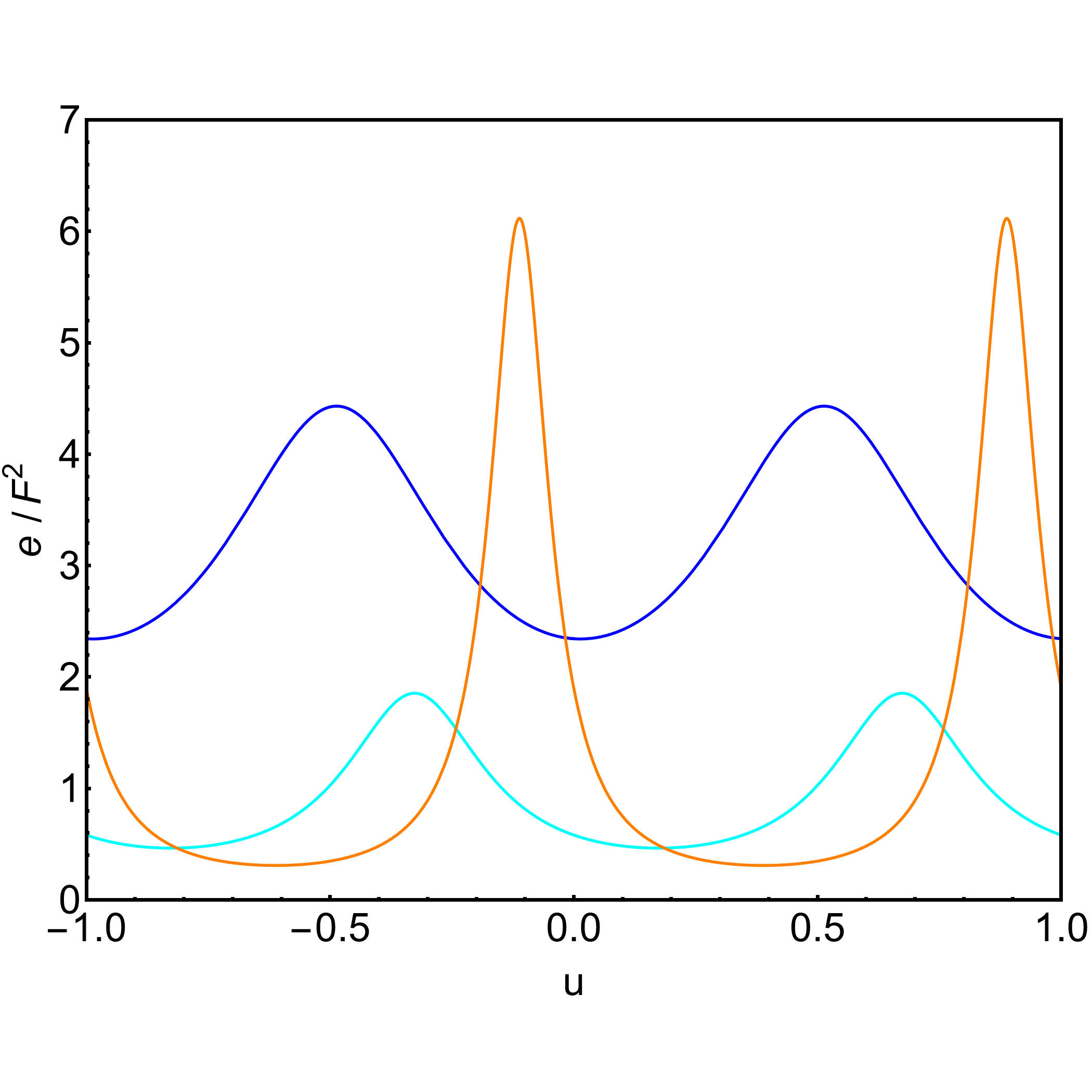}
\captionof{figure} {Behavior of the energy per-oscillator. First row:
  $(\gamma_-,\gamma_+)=(1,1)$. Second row:
  $(\gamma_-,\gamma_+)=(1,1/10)$. Left column: energy computed with
  the limiting expressions for $n\rightarrow\infty$. The oscillating
  part is obtained directly using the Green's function expressions
  with $n=40$ and black dotted curves inside the harmonic spectra zone
  are computed used the energy expression for
  $n\rightarrow\infty$. Red dashed lines define the limits of the
  harmonic spectra. Blue, cyan and orange lines indicate the harmonic
  frequencies $\om=1.0478$, $1.41421$ and $2.101$. Right column:
  Scaled energy $\bar e(r,u)$ around the harmonic frequencies $\om=1.0478$ (blue), $1.41421$ (cyan) and $2.101$ (orange).\label{en7}}         
\end{center}


\section*{Declarations}

\subsection*{Data Availability} Data sharing is not applicable to this article as no datasets were generated or analysed during
  the current study.
\subsection*{Conflict of interest} In addition, the authors have no conflicts of interest to declare that are relevant to the
  content of this article.

\bibliographystyle{amsalpha}

\begin{thebibliography}{A}

  

\bibitem{bo1} C. Bernardin, S. Olla,  {\em Transport Properties of a Chain
  of Anharmonic Oscillators with Random Flip of Velocities},  J. Stat Phys (2011) 145:1224-1255 DOI 10.1007/s10955-011-0385-6

\bibitem{bll} F. Bonetto, J. L. Lebowitz and J.
  Lukkarinen  {\em Fourier’s Law for a Harmonic Crystal with Self-Consistent Stochastic Reservoirs.}
J. of Stat. Physics, Vol. 116,  2004  



   
   
\bibitem{joel57}
  J.L. Lebowitz, P.G. Bergmann, \emph{Irreversible Gibbsian Ensembles},
  Annals of Physics, Vol. 1, N.1, 1-23,
  1957. https://doi.org/10.1016/0003-4916(57)90002-7



\bibitem{carmona} P. Carmona,
  \emph{Existence and uniqueness of an invariant measure
      for a chain of oscillators in contact with two heat baths},
    Stochastic Processes and their Applications 117, (2007), no. 8,
    1076–1092.


\bibitem{evans}     Evans, L. C., 
{\em Weak convergence methods for nonlinear partial differential equations.}
CBMS Regional Conf. Ser. in Math., 74, by the American Mathematical Society, Providence, RI, 1990.

\bibitem{GR} I.S. Gradshteyn and I.M. Ryzhik, {\em Table of Integrals, Series, and Products},
Seventh Edition, Academic Press-Elsevier 2007.
    
    
  


  

\bibitem{klo1} T. Komorowski, J.L. Lebowitz, S. Olla,
  \emph{Heat flow in a periodically forced, thermostatted chain},
  Comm.Math.Phys., 400, 2181–2225 (2023),
  https://doi.org/10.1007/s00220- 023-04654-4




\bibitem{Nak70}   H. Nakazawa 
{\em On the Lattice Thermal Conduction},
Supplement of the Progress of Theoretical Physics, No. 45, (1970), pp 231-262.

\bibitem{prem} Abhinav Prem, Vir B. Bulchandani, and S. L. Sondhi,
  \emph{Dynamics and transport in the boundary-driven dissipative Klein-Gordon chain}, 
Phys. Rev. B 107, 104304, 2023

\bibitem{RLL67}   Rieder, Z., Lebowitz, J.L., Lieb, E.: \emph{Properties of
  harmonic crystal in a stationary non-equilibrium state}. J. Math. Phys. 8, 1073–1078 (1967).


\bibitem{Yag2017}   Yaghoubi M., Foulaadvand M. E., B\'erut, A., \L
  uczka, J., {\em Energetics of a driven Brownian
  harmonic oscillator}, J. Stat. Mech. (2017) 113206  
  
\end{thebibliography}

\end{document}